\documentclass[oneside,AMA,STIX1COL,doublespace,review]{WileyNJD-v2}
\usepackage[T1]{fontenc}
\usepackage[latin9]{inputenc}
\usepackage{array}
\usepackage{float}
\usepackage{booktabs}
\usepackage{url}
\usepackage{multirow}
\usepackage{amsmath}
\usepackage{amssymb}
\usepackage{graphicx}
\usepackage{wasysym}

\makeatletter

\providecommand{\tabularnewline}{\\}
\floatstyle{ruled}
\newfloat{algorithm}{tbp}{loa}
\providecommand{\algorithmname}{Algorithm}
\floatname{algorithm}{\protect\algorithmname}

\articletype{Research Article}%
\address{\orgdiv{Department of Applied Mathematics \& Statistics}, \orgdiv{Institute for Advanced Computational Science}, \orgname{Stony Brook University}, \orgaddress{\city{Stony Brook}, \state{New York}, \country{USA}}}
\corres{Xiangmin Jiao, Department of Applied Mathematics \& Statistics, Institute for Advanced Computational Science, Stony Brook University, Stony Brook, New York, USA. \email{xiangmin.jiao@stonybrook.edu}}
\usepackage{stix}

\@ifundefined{showcaptionsetup}{}{%
 \PassOptionsToPackage{caption=false}{subfig}}
\usepackage{subfig}
\makeatother

\begin{document}
\title{Robust and Efficient Multilevel-ILU Preconditioning of Hybrid \\
Newton-GMRES for Incompressible Navier-Stokes Equations}
\author{Qiao Chen}
\author{Xiangmin Jiao{*}}
\author{Oliver Yang}
\abstract[Summary]{We introduce a robust and efficient preconditioner for a hybrid Newton-GMRES
method for solving the nonlinear systems arising from incompressible
Navier-Stokes equations. When the Reynolds number is relatively high,
these systems often involve millions of degrees of freedom (DOFs),
and the nonlinear systems are difficult to converge, partially due
to the strong asymmetry of the system and the saddle-point structure.
In this work, we propose to alleviate these issues by leveraging a
multilevel ILU preconditioner called \emph{HILUCSI}, which is particularly
effective for saddle-point problems and can enable robust and rapid
convergence of the inner iterations in Newton-GMRES. We further use
Picard iterations with the Oseen systems to hot-start Newton-GMRES
to achieve global convergence, also preconditioned using HILUCSI.
To further improve efficiency and robustness, we use the Oseen operators
as physics-based sparsifiers when building preconditioners for Newton
iterations and introduce adaptive refactorization and iterative refinement
in HILUCSI. We refer to the resulting preconditioned hybrid Newton-GMRES
as \emph{HILUNG}. We demonstrate the effectiveness of HILUNG by solving
the standard 2D driven-cavity problem with Re 5000 and a 3D flow-over-cylinder
problem with low viscosity. We compare HILUNG with some state-of-the-art
customized preconditioners for INS, including two variants of augmented
Lagrangian preconditioners and two physics-based preconditioners,
as well as some general-purpose approximate-factorization techniques.
Our comparison shows that HILUNG is much more robust for solving high-Re
problems and it is also more efficient in both memory and runtime
for moderate-Re problems.}
\maketitle

\section{Introduction\label{sec:Introduction}}

Incompressible Navier-Stokes\emph{ }(INS) equations are widely used
for modeling fluids. The time-dependent INS equations (after normalizing
density) read
\begin{align}
\partial\boldsymbol{u}/\partial t-\nu\Delta\boldsymbol{u}+\boldsymbol{u}\cdot\boldsymbol{\nabla u}+\boldsymbol{\nabla}p & =\boldsymbol{g},\label{eq:momentum}\\
\boldsymbol{\nabla}\cdot\boldsymbol{u} & =0,\label{eq:div-free}
\end{align}
where $\boldsymbol{u}$ and $p$ are velocities and pressure, respectively,
and $\nu$ is the kinetic viscosity. These equations can be solved
using a semi-implicit or fully implicit scheme.\cite{turek1996comparative}
A fully implicit method can potentially enable larger time steps,
but it often leads to large-scale nonlinear systems of equations,
of which robust and efficient solution has been an active research
topic in the past two decades.\cite{elman2014finite,bootland2019preconditioners,pearson2020preconditioners,pernice2001multigrid}
A main challenge in a fully implicit method is to solve the stationary
or quasi-steady INS equation, in which the momentum equation \eqref{eq:momentum}
becomes
\begin{equation}
-\nu\Delta\boldsymbol{u}+\boldsymbol{u}\cdot\boldsymbol{\nabla u}+\boldsymbol{\nabla}p=\boldsymbol{g},\label{eq:steady-momentum}
\end{equation}
which is mathematically equivalent to \eqref{eq:momentum} as the
time step approaches infinity. In this work, we focus on solving the
stationary INS equations. A standard technique to solve this nonlinear
system is to use some variants of \emph{truncated} (aka \emph{inexact})
\emph{Newton methods},\cite[p.284]{heath2018scientific} which solve
the linearized problem approximately at each step, for example, using
an iterative method.\cite{brown1994convergence} Assume INS equations
are discretized using finite elements, such as using the Taylor-Hood
finite-element spaces.\cite{taylor1973numerical} At each truncated
Newton's step, one needs to approximately solve a linear system
\begin{equation}
\begin{bmatrix}\boldsymbol{A} & \boldsymbol{E}^{T}\\
\boldsymbol{E} & \boldsymbol{0}
\end{bmatrix}\begin{bmatrix}\delta\boldsymbol{u}_{k}\\
\delta p_{k}
\end{bmatrix}\approx-\begin{bmatrix}\boldsymbol{f}_{k}\\
\boldsymbol{h}_{k}
\end{bmatrix},\label{eq:nt-sparse}
\end{equation}
where $\delta\boldsymbol{u}_{k}$ and $\delta p_{k}$ correspond to
the increments of $\boldsymbol{u}$ and $p$, respectively; see, e.g.,
Elman et al.\cite{elman2014finite} for a detailed derivation. Let
\[
\boldsymbol{A}=\boldsymbol{K}+\boldsymbol{C}_{k}+\boldsymbol{W}_{k},
\]
where $\boldsymbol{K}$, $\boldsymbol{C}_{k}$, and $\boldsymbol{W}_{k}$
correspond to $\nu\Delta\boldsymbol{u}$, $\boldsymbol{u}_{k}\cdot\boldsymbol{\nabla u}$,
and $\boldsymbol{u}\cdot\boldsymbol{\nabla u}_{k}$, respectively.
In a so-called \emph{hybrid nonlinear method},\cite{heath2018scientific,kelley1995iterative}
inexact Newton methods may be ``hot-started'' using more robust
but more slowly converging methods, such as some Picard iterations.
In the context of INS, a particularly effective hot starter is the
Picard iteration with the Oseen system,\cite{elman2014finite} which
solve the simplified and sparser linear system
\begin{equation}
\begin{bmatrix}\boldsymbol{K}+\boldsymbol{C}_{k} & \boldsymbol{E}^{T}\\
\boldsymbol{E} & \boldsymbol{0}
\end{bmatrix}\begin{bmatrix}\delta\boldsymbol{u}_{k}\\
\delta p_{k}
\end{bmatrix}\approx-\begin{bmatrix}\boldsymbol{f}_{k}\\
\boldsymbol{h}_{k}
\end{bmatrix}.\label{eq:oseen-sparse}
\end{equation}
In the linear-algebra literature,\cite{benzi2005numerical} Eqs.~\eqref{eq:nt-sparse}
and \eqref{eq:oseen-sparse} are often referred to as (\emph{generalized})\emph{
saddle-point problems}, which are notoriously difficult to solve robustly
and efficiently at a large scale. One of the bottlenecks in solving
these nonlinear systems, both in terms of robustness and efficiency,
is the iterative solver of the linear systems at each nonlinear step.
The aim of this paper is to overcome this bottleneck by developing
a robust and efficient preconditioner for those iterative linear solvers
in the inner iterations of a truncated Newton method with hot start
for solving the INS.

For large-scale systems of nonlinear equations, a successful class
of truncated Newton methods is the \emph{Newton-Krylov methods}\cite{brown1990hybrid}
(including\emph{ Jacobian-free Newton-Krylov }methods\cite{knoll2004jacobian,qin2000matrix}),
which utilizes Krylov subspace\emph{ }methods (such as GMRES\cite{saad2003iterative})
to approximate the linear solve. Implementations of such methods can
be found in some general-purpose nonlinear solver libraries, such
as NITSOL,\cite{pernice1998nitsol} MOOSE,\cite{gaston2009moose}
and SNES\cite{brune2015composing} in PETSc.\cite{balay2019petsc}
However, the INS equations pose significant challenges when the Reynolds
number (i.e., $\text{Re}\equiv\Vert\boldsymbol{u}\Vert L/\nu$ with
respect to some reference length $L$) is high, due to steep boundary
layers and potential corner singularities.\cite{turek1996comparative,ghia1982high}
Although one may improve robustness using some generic techniques
such as damping (aka backtracking),\cite{kelley1995iterative} they
often fail for INS.\cite{tuminaro2002backtracking} In recent years,
preconditioners have been recognized as critical techniques in improving
the robustness and efficiency of nonlinear INS solvers. Some of the
most successful preconditioners include (block) incomplete LU,\cite{persson2008newton,ur2008comparison}
augmented Lagrangian methods,\cite{benzi2006augmented,benzi2011modified,farrell2019augmented,moulin2019augmented}
and block preconditioners with approximate Schur complements.\cite{ur2008comparison,elman2006block}
They have been shown to be effective for INS equations with moderate
Re (e.g., up to $2000$)\cite{ur2008comparison,elman2006block}, regularized
flows,\cite{farrel2019alcode} or compressible and Reynolds averaged
Navier-Stokes (RANS) equations with a wide range of Re,\cite{persson2008newton}
but challenges remained for INS with higher Re (see Section~\ref{subsec:2D-drive-cavity-problem}).
In addition, higher Re also requires finer meshes, which lead to larger-scale
systems with millions and even billions of degrees of freedom (DOFs),\cite{lee2015direct}
posing significant challenges in the \textit{\emph{scalability}}\emph{
}of the preconditioners with respect to the problem size.

To address these challenges, we propose a new type of preconditioner
for Newton-GMRES for the INS, based on a multilevel incomplete LU
(MLILU) technique. We build our preconditioner based on \emph{HILUCSI}
(or \emph{Hierarchical Incomplete LU-Crout with Scalability-oriented
and Inverse-based dropping}), which the authors and co-workers introduced
recently for indefinite linear systems from partial differential equations
(PDEs), such as saddle-point problems.\cite{chen2021hilucsi} In this
work, we incorporate HILUCSI into Newton-GMRES to develop \emph{HILUNG},
for nonlinear saddle-point problems from Navier-Stokes equations.
To this end, we introduce sparsifying operators based on \eqref{eq:nt-sparse}
and \eqref{eq:oseen-sparse}, develop adaptive refactorization and
thresholding to avoid potential \textquotedblleft over-factorization\textquotedblright{}
(i.e., too dense incomplete factorization or too frequent refactorization),
and introduce iterative refinement during preconditioning to reduce
memory requirement. As a result, HILUNG can robustly solve the standard
(instead of the regularized) 2D driven-cavity problem with Re $5000$
without stabilization or regularization. In contrast, the state-of-the-art
block preconditioner based on approximate Schur complements\cite{ers14,ifiss}
and (modified) augmented Lagrangian methods\cite{benzi2006augmented,benzi2011modified,farrell2019augmented,moulin2019augmented}
failed to converge at Re $1000$ and/or Re $5000$ with a similar
configuration, respectively. In addition, we show that HILUNG also
improved the efficiency over another general-purpose multilevel ILU
preconditioner\cite{bollhofer2011ilupack} by a factor of 34 for the
3D flow-over-cylinder problem with one million DOFs, and enabled an
efficient solution of the problem with about ten million DOFs using
only 60GB of RAM, while other alternatives ran out of memory. We have
released the core component of HILUNG, namely HILUCSI, as an open-source
library at \url{https://github.com/hifirworks/hifir}.

The remainder of the paper is organized as follows. Section~\ref{sec:Background}
reviews some background on truncated and inexact Newton methods and
preconditioning techniques, especially variants of augmented-Lagrangian,
approximate-Schur-complement, and incomplete-LU preconditioners. In
Section~\ref{sec:MethSec}, we describe the overall algorithm of
HILUNG and its core components for achieving robustness and efficiency.
In Section~\ref{sec:Numerical-results}, we present comparison results
of HILUNG with some state-of-the-art preconditioners and software
implementations. Finally, Section~\ref{sec:Conclusions} concludes
the paper with a discussion on future work.

\section{Background}

\label{sec:Background}In this section, we present some preliminaries
for this work, namely inexact or truncated Newton methods enhanced
by ``hot start'' to achieve global convergence and safeguarded by
damping for robustness. We then review some state-of-the-art preconditioning
techniques for INS, especially those based on approximate Schur complements,
augmented Lagrangian, incomplete LU, and multilevel methods.

\subsection{Preliminary: inexact Newton with hot start and damping\label{subsec:Inexact-Newton}}

Given a system of nonlinear equations $\boldsymbol{F}(\boldsymbol{x})=\boldsymbol{0}$,
where $\boldsymbol{F}:\mathbb{R}^{n}\rightarrow\mathbb{R}^{n}$ is
a nonlinear mapping, let $\boldsymbol{J}(\boldsymbol{x})=[\partial F_{i}/\partial x_{j}]_{ij}$
be its Jacobian matrix. Starting from an initial solution $\boldsymbol{x}_{0}$,
Newton's method (aka the Newton-Raphson method) iteratively seeks
approximations $\boldsymbol{x}_{k+1}=\boldsymbol{x}_{k}+\boldsymbol{s}_{k}$
until the relative residual is sufficiently small, i.e.,
\begin{equation}
\left\Vert \boldsymbol{F}(\boldsymbol{x}_{k})\right\Vert \le\sigma\left\Vert \boldsymbol{F}(\boldsymbol{x}_{0})\right\Vert .\label{eq:term-nt}
\end{equation}
The increment $\boldsymbol{s}_{k}$ is the solution of $\boldsymbol{J}(\boldsymbol{x}_{k})\boldsymbol{s}_{k}=-\boldsymbol{F}(\boldsymbol{x}_{k})$.
In general, $\boldsymbol{s}_{k}$ only needs to be solved approximately
so that 
\begin{equation}
\left\Vert \boldsymbol{J}\left(\boldsymbol{x}_{k}\right)\boldsymbol{s}_{k}+\boldsymbol{F}\left(\boldsymbol{x}_{k}\right)\right\Vert \le\eta_{k}\left\Vert \boldsymbol{F}\left(\boldsymbol{x}_{k}\right)\right\Vert ,\label{eq:in}
\end{equation}
where $\eta_{k}\in[0,\eta_{\max}]$ is the ``forcing parameter.''\cite{dembo1982inexact,eisenstat1994globally}
When $\eta_{k}>0$, the method is known as \emph{inexact Newton}.\cite{dembo1982inexact}
A carefully chosen $\eta_{k}$ preserves the quadratic convergence
of Newton's method when $\boldsymbol{x}_{k}$ is close enough to the
true solution $\boldsymbol{x}_{*}$ while being more efficient when
$\boldsymbol{x}_{k}$ is far from $\boldsymbol{x}_{*}$.\cite{eisenstat1996choosing}
Solving $\boldsymbol{s}_{k}$ beyond the optimal $\eta_{k}$ is called
``over-solving,'' which incurs unnecessary cost and may even undermine
robustness.\cite{brown1990hybrid,eisenstat1996choosing,kelley1995iterative}
For efficiency, iterative methods, such as Krylov subspace methods,
are often used to solve $\boldsymbol{J}(\boldsymbol{x}_{k})\boldsymbol{s}_{k}=-\boldsymbol{F}(\boldsymbol{x}_{k})$
in the inner steps, leading to the so-called \emph{Newton-Krylov methods}
for solving $\boldsymbol{F}(\boldsymbol{x})=\boldsymbol{0}$, such
as the \emph{Newton-GMRES methods}.\cite{brown1990hybrid}

Both exact and inexact Newton methods may fail to converge if the
initial solution is too far from the true solution $\boldsymbol{x}_{*}$.
To improve robustness, \emph{damped Newton}\cite{heath2018scientific}
or \emph{inexact Newton with backtracking}\cite{dennis1996numerical}
introduces a damping (or line search) factor $\omega$ to the increment
(aka direction) $\boldsymbol{s}_{k}$, i.e.,
\begin{equation}
\boldsymbol{x}_{k+1}=\boldsymbol{x}_{k}+\omega\boldsymbol{s}_{k},\label{eq:damp-nt}
\end{equation}
so that $\boldsymbol{x}_{k+1}$ decreases the residual, i.e., $\left\Vert \boldsymbol{F}\left(\boldsymbol{x}_{k+1}\right)\right\Vert <\left\Vert \boldsymbol{F}\left(\boldsymbol{x}_{k}\right)\right\Vert $.
Global convergence may be achieved by using a more robust but more
slowly converging method to ``hot start'' Newton. In this context
of INS, Picard iterations with the Oseen systems as given in \eqref{eq:oseen-sparse}
are particularly effective to hot-start Newton owing to their global
convergence properties under some reasonable assumptions.\cite{elman2014finite}
It is also more efficient to solve the Oseen system at each Picard
iteration in that the Oseen operator is sparser than the Jacobian
matrix in Newton's method. In this work, we use the hybrid Newton-GMRES
with Oseen systems as a physics-based hot starter with some simple
damping as a safeguard. We will focus on preconditioning the linear
systems in this hybrid Newton-GMRES method. 

Although our focus is on the linear solvers, we briefly review some
other nonlinear aspects to motivate our choice. Besides the aforementioned
inexact Newton and backtracking techniques,\cite{dembo1982inexact,eisenstat1994globally}
 more sophisticated general-purpose backtracking techniques have
been proposed.\cite{an2007globally,bellavia2001globally} For example,
Bellavia and Morini\cite{bellavia2001globally} developed Newton-GMRES
with backtracking (NGB), which computes a damping factor $\omega$
iteratively as
\begin{align}
\omega\in\left[\omega_{l},\omega_{u}\right]\text{ and }\eta_{k}=1-\omega\left(1-\eta_{k}\right)\qquad\text{with}\qquad\left\Vert \boldsymbol{F}\left(\boldsymbol{x}_{k}+\omega\boldsymbol{s}_{k}\right)\right\Vert  & \le\left(1-t(1-\eta_{k})\right)\left\Vert \boldsymbol{F}\left(\boldsymbol{x}_{k}\right)\right\Vert ,\label{eq:gin-backtracking}
\end{align}
where $t\in(0,1)$ and $0<\omega_{l}<\omega_{u}<1$. If NGB fails
to converge with such a backtracking strategy, Bellavia and Morini
proposed to switch to a more sophisticated strategy called \emph{equality
curve backtracking} (\emph{ECB}), which constructs an alternative
direction such that \eqref{eq:in} holds the equality. An and Bai\cite{an2007globally}
replaced ECB with an even more sophisticated strategy called \emph{quasi-conjugate-gradient
backtracking} (\emph{QCGB}), which finds a direction in $\text{span}\{\boldsymbol{s}_{k-1},\tilde{\boldsymbol{g}}_{k}\}$,
where $\tilde{\boldsymbol{g}}_{k}$ is the orthogonal projection of
the gradient $\boldsymbol{g}_{k}=\boldsymbol{J}(\boldsymbol{x}_{k})^{T}\boldsymbol{F}(\boldsymbol{x}_{k})$
onto the Krylov subspace.  In this work, we found that when leveraging
Oseen systems as a hot starter for Newton-GMRES in the context of
INS, backtracking is rarely needed, and hence a simple strategy suffices.
More importantly, the more sophisticated backtracking techniques (such
as ECB and QCGB) also require the inner iterations to be solved reliably
with a robust preconditioner.\cite{an2007globally,bellavia2001globally}
Hence, even though we focus on the preconditioning issue for the hybrid
Newton-GMRES with a simple damping strategy, our proposed preconditioning
technique may also benefit Newton-GMRES with more sophisticated backtracking
strategies for other nonlinear equations.

\subsection{Block triangular approximate Schur complements\label{subsec:Schur-complement}}

For INS equations, the resulting systems have a saddle-point structure
(see, e.g., Eqs.~\eqref{eq:nt-sparse}and~\eqref{eq:oseen-sparse}).
A family of ``physics-based'' preconditioners can be derived based
on the block triangular operator
\begin{equation}
\boldsymbol{T}=\begin{bmatrix}\boldsymbol{A} & \boldsymbol{E}^{T}\\
\boldsymbol{0} & \boldsymbol{S}_{T}
\end{bmatrix},\label{eq:schur-tri-op}
\end{equation}
where $\boldsymbol{A}$ is defined as in \eqref{eq:nt-sparse}, and
$\boldsymbol{S}_{T}=-\boldsymbol{E}\boldsymbol{A}^{-1}\boldsymbol{E}^{T}$
is the \emph{Schur complement}. Let $\mathcal{A}$ denote the block
matrix in \eqref{eq:nt-sparse}. If complete factorization (e.g.,
Gaussian elimination) is used, then using $\boldsymbol{T}$ as a preconditioner
of $\mathcal{A}$ enables a Krylov subspace method to converge in
two iterations,\cite{murphy2000note} compared to one iteration when
using $\mathcal{A}$ itself as the preconditioner under exact arithmetic.
Different approximations of $\boldsymbol{S}_{T}$ lead to different
preconditioners.  Most notably, the \emph{pressure convection diffusion
}(\emph{PCD})\cite{silvester2001efficient,kay2002preconditioner}
approximates the Schur complement by
\begin{equation}
\boldsymbol{S}_{T}=-\boldsymbol{E}\boldsymbol{A}^{-1}\boldsymbol{E}^{T}\approx-\boldsymbol{K}_{p}\boldsymbol{F}_{p}^{-1}\boldsymbol{M}_{p},\label{eq:pcd}
\end{equation}
where $\boldsymbol{K}_{p}$ is the pressure Laplacian matrix, $\boldsymbol{F}_{p}$
is a discrete convection-diffusion operator on the pressure space,
and $\boldsymbol{M}_{p}$ is the pressure mass matrix.  The \emph{least-squares
commutator} (\emph{LSC})\cite{elman2006block} approximates the Schur
complement by
\begin{equation}
\boldsymbol{S}_{T}=-\boldsymbol{E}\boldsymbol{A}^{-1}\boldsymbol{E}^{T}\approx-\left(\boldsymbol{E}\boldsymbol{M}_{\boldsymbol{u}}^{-1}\boldsymbol{E}^{T}\right)\left(\boldsymbol{E}\boldsymbol{M}_{\boldsymbol{u}}^{-1}\boldsymbol{A}\boldsymbol{M}_{\boldsymbol{u}}^{-1}\boldsymbol{E}^{T}\right)^{-1}\left(\boldsymbol{E}\boldsymbol{M}_{\boldsymbol{u}}^{-1}\boldsymbol{E}^{T}\right),\label{eq:lsc}
\end{equation}
where $\boldsymbol{M}_{\boldsymbol{u}}$ is the velocity mass matrix.
 Special care is required when imposing boundary conditions.  The
implementations of PCD and LSC often use complete factorization for
its subdomains for smaller systems.\cite{elman2014finite,ifiss} For
large-scale problems, some variants of ILUs or iterative techniques
may be used to approximate $\boldsymbol{A}^{-1}$ in \eqref{eq:schur-tri-op},
$\boldsymbol{K}_{p}^{-1}$ and $\boldsymbol{M}_{p}^{-1}$ in \eqref{eq:pcd},
and $\boldsymbol{M}_{\boldsymbol{u}}^{-1}$ in \eqref{eq:lsc}. We
refer readers to Elman et al.\cite{elman2014finite} for more details
and ur Rehman et al.\cite{ur2008comparison} for some comparisons.

PCD and LSC can be classified accurately as block upper triangular
approximate Schur complement preconditioners. For brevity, we will
refer to them as \emph{approximate Schur complements}. These methods
have been successfully applied to preconditioning laminar flows for
some applications (such as Re 100 in Bootland et al.\cite{bootland2019preconditioners}).
However, these preconditioners are not robust for relatively high
Reynolds numbers (see Section\ \ref{subsec:2D-drive-cavity-problem}).
The lack of robustness is probably because these preconditioners construct
$\boldsymbol{M}$ to approximate $\boldsymbol{T}$, which are suboptimal
compared to preconditioners that construct $\boldsymbol{M}^{-1}$
to approximate $\mathcal{A}^{-1}$ accurately.

\subsection{\label{subsec:Augmented-Lagrangian-preconditio}Augmented Lagrangian
preconditioners}

Benzi and Olshanskii\cite{benzi2006augmented} introduced a preconditioning
technique for the Oseen equations based on the augmented Lagrangian
(AL) method, which was recently adopted by Farrell et al. for the
INS.\cite{farrell2019augmented} We follow the description for the
INS of Farrell et al.\cite{farrell2019augmented} Instead of solving
\eqref{eq:steady-momentum}, the AL-based technique solves a modified
momentum equation
\begin{equation}
-\nu\Delta\boldsymbol{u}+\boldsymbol{u}\cdot\boldsymbol{\nabla u}+\boldsymbol{\nabla}p-\gamma\boldsymbol{\nabla}\boldsymbol{\nabla}\cdot\boldsymbol{u}=\boldsymbol{g}.\label{eq:al-momentum}
\end{equation}
The grad-div term $\gamma\boldsymbol{\nabla}\boldsymbol{\nabla}\cdot\boldsymbol{u}$
is analogous to the penalty term in the augmented Lagrangian method
for optimization problems,\cite{fortin2000augmented} and hence the
name ``AL.''  At the $k$th Newton's step, the weak form of \eqref{eq:steady-momentum}
leads to a linear system 
\begin{equation}
\begin{bmatrix}\boldsymbol{A}+\gamma\boldsymbol{E}^{T}\boldsymbol{M}_{p}^{-1}\boldsymbol{E} & \boldsymbol{E}^{T}\\
\boldsymbol{E} & \boldsymbol{0}
\end{bmatrix}\begin{bmatrix}\delta\boldsymbol{u}_{k}\\
\delta p_{k}
\end{bmatrix}\approx-\begin{bmatrix}\boldsymbol{f}_{k}+\gamma\boldsymbol{E}^{T}\boldsymbol{M}_{p}^{-1}\boldsymbol{h}_{k}\\
\boldsymbol{h}_{k}
\end{bmatrix},\label{eq:linear-system-augmented}
\end{equation}
where $\boldsymbol{M}_{p}$ is the mass matrix for the pressure. Benzi
and Olshanskii\cite{benzi2006augmented} argued that for sufficiently
large $\gamma$, the Schur complement can be approximated reasonably
well by 
\begin{equation}
\boldsymbol{S}^{-1}\approx-(\nu+\gamma)\boldsymbol{M}_{p}^{-1}.\label{eq:approximate-Schur-complement}
\end{equation}
It is worth noting that their analysis assumed that the modified momentum
equation was not only used as the preconditioner but also as the primary
discretization of the INS. A major challenge is to solve the leading
block efficiently. Specialized multigrid methods can be employed to
overcome this challenge.\cite{benzi2006augmented,farrell2019augmented}
The use of multigrid methods also makes the preconditioner relatively
insensitive to the mesh resolution.

To improve the generality of the AL method, Benzi, Olshanskii, and
Wang\cite{benzi2011modified} proposed a simplification of the AL
by approximating the leading block $\boldsymbol{A}+\gamma\boldsymbol{E}^{T}\boldsymbol{M}_{p}^{-1}\boldsymbol{E}$
in \eqref{eq:linear-system-augmented} with a block upper-triangular
matrix and then modifying the approximation to the Schur complement
accordingly. Several variants of the approximate Schur complement
had been developed in the literature,\cite{benzi2011modified,he2020efficient}
and they are collectively referred to as the modified AL (MAL). A
subtle issue in both AL and MAL is the choice of $\gamma$: As in
other AL methods, too large $\gamma$ in AL and MAL may lead to ill-conditioning,\cite{nocedal2006numerical}
but a small $\gamma$ leads to inaccurate approximation to the Schur
complement. Farrell et al.\cite{farrell2019augmented} suggested using
$\gamma=10^{4}$. For MAL, because the dropping of the lower-triangular
block suggests that $\gamma$ cannot be too large,\cite{benzi2011modified}
but a small $\gamma$ implies difficulties in constructing an accurate
approximation to the Schur complement. Moulin et al.\cite{moulin2019augmented}
suggested using $\gamma$ between $0.1$ and $1$. In Section~\ref{sec:Numerical-results},
we will compare our proposed approach with the recent AL implementation
of Farrell et al.\cite{farrell2019augmented} and MAL implementation
of Moulin et al.\cite{moulin2019augmented}

\subsection{Single-level and multilevel ILUs\label{subsec:Multilevel-ILU}}

\emph{Incomplete LU }(\emph{ILU}) is arguably one of the most successful
general preconditioning techniques for Krylov subspace methods. Given
a linear system $\boldsymbol{A}\boldsymbol{x}=\boldsymbol{b}$, ILU
approximately factorizes $\boldsymbol{A}$ by
\begin{equation}
\boldsymbol{P}^{T}\boldsymbol{AQ}\approx\boldsymbol{L}\boldsymbol{D}\boldsymbol{U},\label{eq:single-level-ILU}
\end{equation}
where $\boldsymbol{D}$ is a diagonal matrix, and $\boldsymbol{L}$
and $\boldsymbol{U}$ are unit lower and upper triangular matrices,
respectively. The permutation matrices $\boldsymbol{P}$ and $\boldsymbol{Q}$
may be constructed statically (such as using equilibration\cite{duff2001algorithms}
or reordering\cite{amestoy1996approximate}) and dynamically (such
as by pivoting\cite{saad1988preconditioning,saad2003iterative}).
We refer to \eqref{eq:single-level-ILU} as \emph{single-level ILU}.
The simplest form of ILU is ILU($0$), which does not have any pivoting
and preserves the sparsity patterns of the lower and upper triangular
parts of $\boldsymbol{P}^{T}\boldsymbol{AQ}$ in $\boldsymbol{L}$
and $\boldsymbol{U}$, respectively. To improve the effectiveness
of ILU, one may introduce \emph{fills} (aka \emph{fill-ins}), which
are nonzeros entries in $\boldsymbol{L}$ and $\boldsymbol{U}$ that
do not exist in the sparsity patterns of the lower and upper triangular
parts of $\boldsymbol{P}^{T}\boldsymbol{AQ}$, respectively. The fills
can be introduced based on their levels in the elimination tree or
based on the magnitude of numerical values. The former leads to the
so-called ILU($k$), which zeros out all the fills of level $k+1$
or higher in the elimination tree. It is worth noting that ILU($k$)
(including ILU($0$)) was advocated for preconditioning Navier-Stokes
by several authors in the literature.\cite{persson2008newton,ur2008comparison,yang2014scalable}
\emph{ILU with dual thresholding} (\emph{ILUT})\cite{saad1994ilut}
introduces fills based on both their levels in the elimination tree
and their numerical values. To overcome tiny pivots, one may enable
pivoting, leading to so-called ILUP\cite{saad1988preconditioning}
and ILUTP.\cite{saad2003iterative} However, such approaches cannot
prevent small pivots and may suffer from instabilities.\cite{saad2005multilevel}
Recently, Konshin et al.\cite{konshin2017lu} studied the use of a
variant of ILUT for the INS with low Reynolds numbers and Oseen linearization.

\emph{Multilevel incomplete LU} (\emph{MLILU}) is another general
algebraic framework for building block preconditioners. More precisely,
let $\boldsymbol{A}$ be the input coefficient matrix. A two-level
ILU reads
\begin{equation}
\boldsymbol{P}^{T}\boldsymbol{AQ}=\begin{bmatrix}\boldsymbol{B} & \boldsymbol{F}\\
\boldsymbol{E} & \boldsymbol{C}
\end{bmatrix}\approx\boldsymbol{M}=\begin{bmatrix}\tilde{\boldsymbol{B}} & \tilde{\boldsymbol{F}}\\
\tilde{\boldsymbol{E}} & \boldsymbol{C}
\end{bmatrix}=\begin{bmatrix}\boldsymbol{L}_{B} & \boldsymbol{0}\\
\boldsymbol{L}_{E} & \boldsymbol{I}
\end{bmatrix}\begin{bmatrix}\boldsymbol{D}_{B} & \boldsymbol{0}\\
\boldsymbol{0} & \boldsymbol{S}_{C}
\end{bmatrix}\begin{bmatrix}\boldsymbol{U}_{B} & \boldsymbol{U}_{F}\\
\boldsymbol{0} & \boldsymbol{I}
\end{bmatrix},\label{eq:two-level-ILU}
\end{equation}
where $\boldsymbol{B}\approx\tilde{\boldsymbol{B}}=\boldsymbol{L}_{B}\boldsymbol{D}_{B}\boldsymbol{U}_{B}$
corresponds to a single-level ILU of the leading block, and $\boldsymbol{S}_{C}=\boldsymbol{C}-\boldsymbol{L}_{E}\boldsymbol{D}_{B}\boldsymbol{U}_{F}$
is the Schur complement. Like single-level ILU, the permutation matrices
$\boldsymbol{P}$ and $\boldsymbol{Q}$ can be statically constructed.
One can also apply pivoting\cite{mayer2007multilevel} or deferring\cite{bollhofer2006multilevel,chen2021hilucsi}
in MLILU. For this two-level ILU, $\boldsymbol{P}\boldsymbol{M}\boldsymbol{Q}^{T}$
provides a preconditioner of $\boldsymbol{A}$. By factorizing $\boldsymbol{S}_{C}$
in \eqref{eq:two-level-ILU} recursively with the same technique,
we then obtain a multilevel ILU and a corresponding multilevel preconditioner.
The recursion terminates when the Schur complement is sufficiently
small, and then a complete factorization (such as LU with partial
pivoting) can be employed. Compared to single-level ILUs, MLILU is
generally more robust and effective for indefinite systems.\cite{ghai2017comparison,chen2021hilucsi}
It is worth noting that MLILU differs from approximate Schur complements\cite{elman2006block,elman2014finite}
and other physics-based block preconditioners (such as SIMPLE\cite{elman2008taxonomy,pernice2001multigrid}):
the blocks in MLILU are constructed algebraically and hence are different
from the block structures obtained from the PDEs (such as those in
\eqref{eq:nt-sparse} and \eqref{eq:oseen-sparse}), and there are
typically more than two levels of blocks. In this work, we utilize
a multilevel ILU technique called HILUCSI,\cite{chen2021hilucsi}
which we will describe in more detail in Section~\ref{subsec:HILUCSI}.

\subsection{Multigrid preconditioners}

Besides MLILU, another popular multilevel approach is the \emph{multigrid
methods}, including \emph{geometric multigrid} (\emph{GMG}),\cite{briggs2000multigrid}\emph{
algebraic} \emph{multigrid} (\emph{AMG}),\cite{briggs2000multigrid}
and their hybrids.\cite{lu2014hybrid,rudi2015extreme} Multigrid methods
are particularly successful in solving elliptic PDEs, such as the
Poisson equation arising from semi-implicit discretizations of INS\cite{ghia1982high,pernice2001multigrid}
or from subdomain problems in approximate-Schur-complement approaches.\cite{elman2003parallel}
However, for saddle-point problems arising from fully implicit discretizations,
the state-of-the-art multigrid methods are less robust than incomplete
LU.\cite{ghai2017comparison} Hence, we do not consider them in this
work besides its use as a component in the AL preconditioner as mentioned
in Section~\ref{subsec:Augmented-Lagrangian-preconditio}.

\section{Achieving robustness and efficiency with HILUNG}

\label{sec:MethSec}
\begin{figure}
\begin{centering}
\includegraphics[width=0.7\columnwidth]{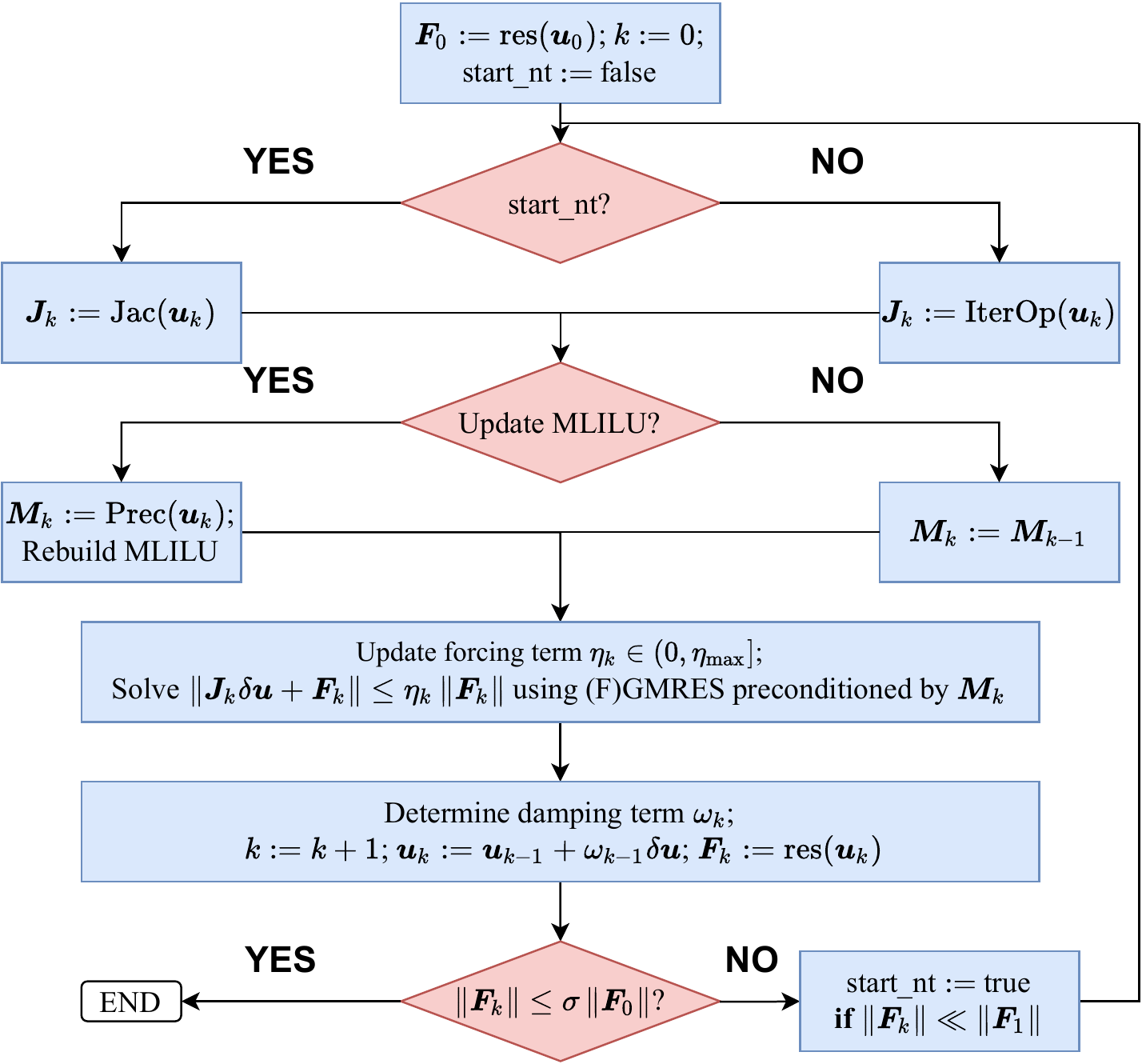}
\par\end{centering}
\caption{\label{fig:HILUNG-algorithm-flowchart}HILUNG algorithm flowchart.}
\end{figure}

We now describe \emph{HILUNG}, or \emph{HILUcsi-preconditioned
Newton-Gmres}. HILUNG uses the hybrid Newton-GMRES as its baseline
nonlinear solver, with Oseen systems for hot start and damping as
a safeguard. Figure~\ref{fig:HILUNG-algorithm-flowchart} illustrates
the overall control flow of HILUNG, which shares some similarities
as others (such as those of Eisenstat and Walker\cite{eisenstat1996choosing}
and of Pernice and Walker\cite{pernice1998nitsol}). However, one
of its core components is the HILUCSI-based preconditioner for the
Oseen systems and Newton iterations. Within each of these nonlinear
steps, HILUNG has three key components: First, determine a suitable
forcing parameter; second, solve the corresponding approximated increments
using preconditioned GMRES; third, apply a proper damping factor
to the increment to safeguard against overshooting. We will describe
these components in more detail below, focusing on HILUCSI and its
theoretical foundation.

\subsection{HILUCSI\label{subsec:HILUCSI}}

The computational kernel of HILUNG is a robust and efficient multilevel
ILU preconditioner, called HILUCSI (or Hierarchical Incomplete LU-Crout
with Scalability-oriented and Inverse-based dropping), which the authors
developed recently.\cite{chen2021hilucsi} HILUCSI shares some similarities
with other MLILU (such as ILUPACK\cite{bollhofer2011ilupack}) in
its use of the Crout version of ILU factorization,\cite{li2003crout}
its dynamic deferring of rows and columns to ensure the well-conditioning
of $\tilde{\boldsymbol{B}}$ in \eqref{eq:two-level-ILU} at each
level,\cite{bollhofer2006multilevel} and its inverse-based dropping
for robustness.\cite{bollhofer2006multilevel} Different from ILUPACK,
however, HILUCSI improved the robustness for saddle-point problems
from PDEs by using static deferring of small diagonals and by utilizing
a combination of symmetric and unsymmetric permutations at the top
and lower levels, respectively. Furthermore, HILUCSI introduced a
\emph{scalability-oriented dropping} to achieve near-linear time complexity
in its factorization and triangular solve. As a result, HILUCSI is
particularly well suited for preconditioning large-scale systems arising
from INS equations. We refer readers to Chen et al.\cite{chen2021hilucsi}
for details of HILUCSI and a comparison with some state-of-the-art
ILU preconditioners (including ILUPACK\cite{bollhofer2011ilupack}
and supernodal ILUTP\cite{li2011supernodal}) for large-scale indefinite
systems.

In the context of preconditioning GMRES for INS, we precondition the
Jacobian matrix $\boldsymbol{J}$ (i.e., $\boldsymbol{A}=\boldsymbol{J}$
in \eqref{eq:single-level-ILU} and \eqref{eq:two-level-ILU}). For
efficiency, we apply HILUCSI on a sparsified version of $\boldsymbol{J}$,
which we denoted by $\boldsymbol{J}_{S}$ and refer to it as the \emph{sparsifying
operator} (or simply \emph{sparsifier}). Within Newton iterations,
the sparsifier may be the Oseen operator utilizing a previous solution
in its linearization. Another potential sparsifier is a lower-order
discretization method (see, e.g., Persson and Peraire\cite{persson2008newton}).
The sparsifier is also related to physics-based preconditioners,\cite{elman2008taxonomy}
except that $\boldsymbol{J}_{S}$ is less restrictive than physics-based
preconditioners and hence is easier to construct. In HILUCSI, we note
two key parameters in HILUCSI: 1) $\alpha$ for scalability-oriented
dropping, which limits the number of nonzeros (nnz) in each column
of $\begin{bmatrix}\boldsymbol{L}_{B}\\
\boldsymbol{L}_{E}
\end{bmatrix}$ and $\boldsymbol{U}_{F}$ and in each row of $\begin{bmatrix}\boldsymbol{U}_{B} & \boldsymbol{U}_{F}\end{bmatrix}$.
2) droptol, which controls inverse-based dropping. In particular,
we limit $\boldsymbol{L}_{E}$ and $\boldsymbol{U}_{F}$ at each level
to be within $\alpha$ times the nnz in the corresponding row and
column of $\boldsymbol{J}$ subject to a safeguard for rows and columns
with a small nnz in $\boldsymbol{J}$. A larger $\alpha$ and a smaller
droptol lead to more accurate but also more costly incomplete factors.
Hence, we need to balance accuracy and efficiency by adapting these
parameters, so that we can achieve robustness while avoiding ``over-factorization''
in HILUCSI. It is also desirable for the approximation error in the
sparsifier (i.e., $\boldsymbol{J}-{\boldsymbol{J}_{S}}$) to be commensurate
with the droppings in HILUCSI.

For INS, there is a connection between HILUCSI and the approximate
Schur complements, such as PCD and LSC described in Section~\ref{subsec:Schur-complement}.
Specifically, HILUCSI defers all small diagonals directly to next
level after applying equilibration,\cite{duff2001algorithms} which
we refer to as \emph{static deferring}. At the first level, the static
deferring is likely recover the saddle-point structure as in \eqref{eq:nt-sparse}
or \eqref{eq:oseen-sparse}. However, HILUCSI constructs a preconditioner
in the form of $\text{\ensuremath{\boldsymbol{M}}}\approx\begin{bmatrix}\tilde{\boldsymbol{B}} & \tilde{\boldsymbol{F}}\\
\tilde{\boldsymbol{E}} & \boldsymbol{C}
\end{bmatrix}=\begin{bmatrix}\boldsymbol{L}_{B} & \boldsymbol{0}\\
\boldsymbol{L}_{E} & \boldsymbol{I}
\end{bmatrix}\begin{bmatrix}\boldsymbol{D}_{B} & \boldsymbol{0}\\
\boldsymbol{0} & \boldsymbol{S}_{C}
\end{bmatrix}\begin{bmatrix}\boldsymbol{U}_{B} & \boldsymbol{U}_{F}\\
\boldsymbol{0} & \boldsymbol{I}
\end{bmatrix}$ instead of $\boldsymbol{T}=\begin{bmatrix}\boldsymbol{B} & \boldsymbol{E}^{T}\\
\boldsymbol{0} & \boldsymbol{S}_{T}
\end{bmatrix}$ as in PCD and LSC. In other words, HILUCSI preserves more information
in the lower-triangular part than approximate Schur complements. In
addition, HILUCSI guarantees that $\tilde{\boldsymbol{B}}$ is well-conditioned
by dynamically deferring rows and columns to the next level, but $\boldsymbol{B}$
may be ill-conditioned in $\boldsymbol{T}$. For these reasons, we
expect HILUCSI to enable faster convergence and deliver better robustness
than PCD and LSC, as we will confirm in Section~\ref{sec:Numerical-results}.
In addition, the implementations of PCD and LSC often rely on complete
factorization for its subdomains,\cite{elman2014finite,ifiss} but
HILUCSI uses incomplete factorization to obtain $\tilde{\boldsymbol{B}}$
and it factorizes $\boldsymbol{S}_{C}$ recursively. Hence, we expect
HILUCSI to deliver better absolute performance per iteration than
PCD and LSC. From practical point of view, HILUCSI is also more user-friendly
than PCD and LSC, in that it is purely algebraic and does not require
the users to modify their PDE codes.

We have previously conducted a thorough analysis of HILUCSI in terms
of its accuracy\cite{jiao2020approximate} and its efficiency.\cite{chen2021hilucsi}
In particular, HILUCSI was shown to satisfy an \emph{$\epsilon$-accuracy}
criterion (Definition 3.8 of Jiao and Chen\cite{jiao2020approximate}).
As a result, with sufficiently small dropping thresholds, HILUCSI
guarantees the convergence of right-preconditioned GMRES, and it converges
to an optimal preconditioner as the thresholds decrease, enabling
the right-preconditioned GMRES to converge in one iteration at the
limit (see Lemma 4.3 and Theorem 3.6 of Jiao and Chen\cite{jiao2020approximate}).
Furthermore, the analysis of $\epsilon$-accuracy also applies to
singular systems, which is relevant to INS since the Jacobian matrix
is singular, with a null space corresponding to the constant mode
of the pressure (aka the hydrostatic pressure). In addition, we proved
that HILUCSI has (near) linear space and time complexities in its
factorization cost and its solve cost per GMRES iteration.\cite{chen2021hilucsi}
In contrast, neither the block triangular approximate Schur complement
preconditioners\cite{elman2014finite} nor the augmented Lagrangian
preconditioners\cite{benzi2006augmented,benzi2011modified} satisfy
the $\epsilon$-accuracy criterion, and the complete factorizations
typically used for their leading blocks\cite{ers14,ifiss,moulin2019augmented}
have superlinear space and time complexity. For these reasons, we
expect HILUCSI to perform well as the preconditioner for INS, as we
will confirm empirically in Section~\ref{sec:Numerical-results}.

\subsection{Frequency of factorization}

To use HILUCSI effectively as preconditioners in Newton-GMRES, we
need to answer two questions: First, how frequently should the sparsifier
be recomputed and factorized? Second, how accurate should the incomplete
factorization be in terms of $\alpha$ and droptol (cf. Section~\ref{subsec:Multilevel-ILU})?
Clearly, more frequent refactorization and more accurate approximate
factorization may improve robustness. However, they may also lower
efficiency because factorization (including incomplete factorization)
is typically far more expensive than triangular solves. In addition,
a more accurate approximate factorization is also denser in general.
It is desirable to achieve robustness while minimizing over-factorization.
Pernice and Walker\cite{pernice1998nitsol} used a fixed refactorization
frequency to show that it is sometimes advantageous to reuse a previous
preconditioner.

Regarding the first question, we recompute and factorize the sparsifier
if 1) the number of GMRES iterations in the previous nonlinear step
exceeded a user-specified threshold $N$, or 2) the increment in the
previous step is greater than some factor of the previous solution
vector. The rationale of the first criterion is that an excessive
number of GMRES iterations indicates the ineffectiveness of the preconditioner,
which is likely due to an outdated sparsifier (assuming the sparsification
process and HILUCSI are both sufficiently accurate). The second criterion
serves as a safeguard against rapid changes in the solution, especially
at the beginning of the nonlinear iterations. Finally, to preserve
the quadratic convergence of Newton's method, we always build a new
sparsifier and preconditioner at the first Newton iteration. For the
second question, we adapt $\alpha$ and droptol based on whether it
is during Picard or Newton iterations. It is desirable to use smaller
$\alpha$ and larger droptol during Picard iterations with the Oseen
systems for better efficiency and use larger $\alpha$ and smaller
droptol for Newton iterations for faster convergence. Based on our
numerical experimentation, for low Re ($<200$), we use $\alpha=2$,
and we set $\text{droptol}=0.02$ and $0.01$ during Oseen and Newton
iterations, respectively. For high Re, we use $\alpha=5$ by default
and set $\text{droptol}=0.01$ and $\text{droptol}=0.001$, respectively.
It is worth noting that in HILUCSI, the scalability-oriented dropping
dominates when droptol is small, and even setting $\text{droptol}$
to $0$ for HILUCSI would not lead to excessive fills or loss of near-linear
complexity. In contrast, some other ILU techniques, such as ILUTP,
would suffer from superlinear complexity and excessive fills when
droptol is too small.

\subsection{\label{subsec:Improving-robustness-with}Improving robustness with
iterative refinement and null-space elimination}

In HILUNG, the sparsification in $\boldsymbol{J}_{S}$, the delay
of refactorization, and the droppings in HILUCSI all introduce inaccuracies
to the preconditioner $\boldsymbol{M}$. To improve robustness, it
may be beneficial to have a built-in correction in $\boldsymbol{M}$.
To do this, we utilize the concept of \emph{iterative refinement},
which is often used in direct solvers for ill-conditioned systems,\cite{Golub13MC}
and it was also used previously by Dahl and Wille\cite{dahl1992ilu}
in conjunction with single-level ILU. With the use of iterative refinement,
we utilize the \emph{flexible GMRES} (\emph{FGMRES}),\cite{saad1993flexible}
which allows inner iterations within the preconditioner. In our experiments,
we found that two inner iterations are enough and can significantly
improve the effectiveness of the preconditioner when a sparsifier
is used.

In addition, note that the Jacobian matrix may be singular, for example,
when the PDE has a pure Neumann boundary condition. We assume the
null space is known and project off the null-space components during
preconditioning. We refer to it as \emph{null-space elimination}.
In particular, let $\boldsymbol{q}$ be composed of an orthonormal
basis of the (right) null space of $\boldsymbol{J}_{k}$. Given a
vector $\boldsymbol{v}$ and an intermediate preconditioner $\hat{\boldsymbol{M}}$
obtained from HILUCSI, we construct an ``implicit'' preconditioner
$\boldsymbol{M}$, which computes $\boldsymbol{z}=\boldsymbol{M}^{+}\boldsymbol{\boldsymbol{v}}$
iteratively starting with $\boldsymbol{z}_{0}=\boldsymbol{0}$ and
then 
\begin{equation}
\boldsymbol{z}_{n}=\boldsymbol{z}_{n-1}+\boldsymbol{\Pi}\hat{\boldsymbol{M}}^{-1}\left(\boldsymbol{\boldsymbol{v}}-\boldsymbol{J}_{k}\boldsymbol{z}_{n-1}\right),\qquad\text{for }n=1,2...,K,\label{eq:iter-refine}
\end{equation}
where $\boldsymbol{\Pi}=\boldsymbol{I}-\boldsymbol{q}\boldsymbol{q}^{T}$.
If $K=1$, the process results in $\boldsymbol{M}^{+}=\boldsymbol{\Pi}\hat{\boldsymbol{M}}^{-1}$.
For large $K$, the process reduces to a stationary iterative solver,
which converges when $\rho(\boldsymbol{I}-\boldsymbol{\Pi}\hat{\boldsymbol{M}}^{-1}\boldsymbol{J}_{k})<1$,
where $\rho$ denotes the spectral radius. In our experiments, we
found that $K=2$ is effective during Newton iterations, which significantly
improves efficiency for high Re without compromising efficiency for
low Re. Notice that the null-space eliminator $\boldsymbol{\Pi}$
is optional for INS with finite element methods because there exists
a constant mode in the pressure with Dirichlet (i.e., fixed velocity)
boundary conditions applied to all walls. Moreover, both Eqs.~\eqref{eq:nt-sparse}
and~\eqref{eq:oseen-sparse} are range-symmetric, i.e., $\mathcal{N}\left(\boldsymbol{J}_{k}\right)=\mathcal{N}\left(\boldsymbol{J}_{k}^{T}\right)$.
Therefore, for \eqref{eq:nt-sparse} and \eqref{eq:oseen-sparse},
we have both
\begin{equation}
\begin{bmatrix}\boldsymbol{K}+\boldsymbol{C}_{k}+\boldsymbol{W}_{k} & \boldsymbol{E}^{T}\\
\boldsymbol{E} & \boldsymbol{0}
\end{bmatrix}\begin{bmatrix}\boldsymbol{0}\\
\boldsymbol{1}
\end{bmatrix}=\boldsymbol{0}\qquad\text{and}\qquad\begin{bmatrix}\boldsymbol{K}+\boldsymbol{C}_{k} & \boldsymbol{E}^{T}\\
\boldsymbol{E} & \boldsymbol{0}
\end{bmatrix}\begin{bmatrix}\boldsymbol{0}\\
\boldsymbol{1}
\end{bmatrix}=\boldsymbol{0},
\end{equation}
which means $\boldsymbol{J}_{k}$ can automatically eliminate the
null-space component arising from INS. Nevertheless, we observe that
such a null-space eliminator can mitigate the effect of round-off
errors and reduce the number of iterations.

\subsection{Overall algorithm\label{subsec:overall-algorithm}}

For completeness, Algorithm\ \ref{alg:hilung} presents the pseudocode
for HILUNG. The first three arguments of the algorithm, namely $\boldsymbol{F}$,
$\boldsymbol{J}$, and $\boldsymbol{x}_{0}$, are similar to typical
Newton-like methods. We assume the initial solution $\boldsymbol{x}_{0}$
is obtained from some linearized problems (such as the Stokes equations
in the context of INS). Unlike a standard nonlinear solver, HILUNG
has a fourth input argument $\boldsymbol{J}_{S}$, which is a callback
function. $\boldsymbol{J}_{S}$ returns a matrix, on which we compute
the HILUCSI preconditioner $\boldsymbol{M}$; see line~\ref{line:HILUCSI}.
To support hot start, HILUNG allows $\boldsymbol{J}$ to return either
the Oseen operator (during hot start) or the Jacobian matrix (after
hot start); see line~\ref{line:iter-op}. The switch from Oseen to
Newton iterations is specified in line\ \ref{line:hot-start}, based
on the current residual relative to the initial residual. Line~\ref{line:rtol-gmres}
corresponds to the determination of the forcing parameter $\eta_{k}$.
During Picard iterations, it is sufficient to use a constant $\eta_{k}$
due to the linear convergence of Picard iterations.\cite{elman2014finite}
In our tests, we fixed $\eta_{k}$ to be $0.3$. For Newton iterations,
we choose $\eta_{k}$ based on the second choice by Eisenstat and
Walker;\cite{eisenstat1996choosing} specifically, $\eta_{k}=\min\left\{ \eta_{\max},0.9\thinspace\left\Vert \boldsymbol{F}(\boldsymbol{x}_{k})\right\Vert ^{2}/\left\Vert \boldsymbol{F}(\boldsymbol{x}_{k-1})\right\Vert ^{2}\right\} $,
which are further restricted to be no smaller than $0.9\thinspace\eta_{k-1}^{2}$
if $0.9\,\eta_{k-1}^{2}>0.1$.\cite{eisenstat1996choosing} To avoid
over-solving in the last Newton step, we safeguarded $\eta_{k}$ to
be no smaller than $0.5\thinspace\sigma\left\Vert \boldsymbol{F}\left(\boldsymbol{x}_{0}\right)\right\Vert /\left\Vert \boldsymbol{F}\left(\boldsymbol{x}_{k}\right)\right\Vert $.\cite{kelley1995iterative}
Regarding the damping factors, we compute $\omega$ using the Armijo
rule by iteratively halving $\omega$, i.e., $\omega_{j}=\omega_{j-1}/2$
for $j=1,2,\ldots,$ with $\omega_{0}\equiv1$,\cite{kelley1995iterative}
as shown between lines~\ref{line:damping-start} and~\ref{line:damping-end}.
\begin{algorithm}
\caption{\label{alg:hilung}$\boldsymbol{x}=\textrm{\textbf{hilung}}\left(\boldsymbol{F},\boldsymbol{J},\boldsymbol{x}_{0},\boldsymbol{J}_{S},\textrm{args}\right)$}

\begin{raggedright}
$\boldsymbol{F},\boldsymbol{J}$: callback functions for computing
residual and Oseen/Jacobian matrix, respectively.
\par\end{raggedright}
\begin{raggedright}
$\boldsymbol{x}_{0}$: initial solution.
\par\end{raggedright}
\begin{raggedright}
$\boldsymbol{J}_{S}$: callback function for computing sparsifying
operator (can be same as $\boldsymbol{J}$).
\par\end{raggedright}
\begin{raggedright}
args: control parameters.
\par\end{raggedright}
\begin{algorithmic}[1]

\State $\sigma,\eta_{\max},\beta,\epsilon,\alpha,\textrm{droptol},m,N,\theta\leftarrow\text{args}$\label{line:pars}\hfill{}\{control
parameters\}

\State $\boldsymbol{s}_{-1}\leftarrow\boldsymbol{1}$; $\boldsymbol{x}_{-1}\leftarrow\boldsymbol{0}$;
$k=0$

\While{$\left\Vert \boldsymbol{F}\left(\boldsymbol{x}_{k}\right)\right\Vert >\sigma\left\Vert \boldsymbol{F}\left(\boldsymbol{x}_{0}\right)\right\Vert $}

\State $\text{started\_nt}\leftarrow\text{\ensuremath{\left\Vert \boldsymbol{F}\left(\boldsymbol{x}_{k}\right)\right\Vert }}\le\beta\left\Vert \boldsymbol{F}\left(\boldsymbol{x}_{0}\right)\right\Vert $\label{line:hot-start}\hfill{}\{hot-started
Newton if the solution is close\}

\State $\boldsymbol{J}_{k}\leftarrow\boldsymbol{J}\left(\boldsymbol{x}_{k},\textrm{started\_nt}\right)$\label{line:iter-op}\hfill{}\{compute
Jacobian/iteration matrix\}

\If{prev. FGMRES iter. count$\ge N$\textbf{ or} $\left\Vert \boldsymbol{s}_{k-1}\right\Vert \ge\epsilon\left\Vert \boldsymbol{x}_{k-1}\right\Vert $\textbf{
or} first Newton iter.}\label{line:refactor}

\State adapt $\alpha$ and $\text{droptol}$ for HILUCSI based on
$\text{started\_nt}$\label{line:thresholding}

\State $\boldsymbol{J}_{S,k}\leftarrow\boldsymbol{J}_{S}\left(\boldsymbol{x}_{k},\text{started\_nt}\right)$;
construct $\boldsymbol{M}$ from $\boldsymbol{J}_{S,k}$ by factorizing
by HILUCSI\label{line:HILUCSI}

\EndIf

\State determine relative tolerance $\eta_{k}\in\left(0,\eta_{\max}\right]$
based on $\text{started\_nt}$ for FGMRES($m$)\label{line:rtol-gmres}

\State use FGMRES($m$) to solve $\boldsymbol{J}_{k}\boldsymbol{M}^{-1}\boldsymbol{t}_{k}\approx-\boldsymbol{F}\left(\boldsymbol{x}_{k}\right)$
s.t. $\left\Vert \boldsymbol{J}_{k}\boldsymbol{s}_{k}+\boldsymbol{F}\left(\boldsymbol{x}_{k}\right)\right\Vert \le\eta_{k}\left\Vert \boldsymbol{F}\left(\boldsymbol{x}_{k}\right)\right\Vert ,\text{ where }\text{\ensuremath{\boldsymbol{s}_{k}=}\ensuremath{\boldsymbol{M}^{-1}\boldsymbol{t}_{k}}}$

\State $\omega\leftarrow1$\label{line:damping-start}\hfill{}\{initial
damping factor\}

\While{$\left\Vert \boldsymbol{F}\left(\boldsymbol{x}_{k}+\boldsymbol{s}_{k}\right)\right\Vert >\left(1-\theta\thinspace\omega\right)\left\Vert \boldsymbol{F}\left(\boldsymbol{x}_{k}\right)\right\Vert $}\hfill{}\{we
set $\theta=10^{-4}$\}

\State $\omega\leftarrow\omega/2$\label{line:damping}\hfill{}\{halve
the damping factor\}

\State $\boldsymbol{s}_{k}\leftarrow\omega\thinspace\boldsymbol{s}_{k}$

\EndWhile\label{line:damping-end}

\State $\boldsymbol{x}\leftarrow\boldsymbol{x}_{k+1}\leftarrow\boldsymbol{x}_{k}+\boldsymbol{s}_{k}$;
$k\leftarrow k+1$\hfill{}\{update solution and counter\}

\EndWhile

\end{algorithmic}
\end{algorithm}

\section{Numerical results and comparisons}

\label{sec:Numerical-results}

In this section, we assess the robustness and efficiency of HILUNG.
To this end, we compare HILUNG with some of the state-of-the-art customized
preconditioners for INS (including two variants of augmented Lagrangian
methods\cite{benzi2006augmented,benzi2011modified,farrel2019alcode,moulin2019augmented}
and two ``physics-based'' preconditioners\cite{ers14,ifiss}) as
well as some general-purpose (approximate) factorization techniques
(including ILUPACK v2.4\cite{bollhofer2011ilupack}, MUMPS v5.3.5\cite{amestoy2000mumps,amestoy2019performance},
and the popular ILU($k$)\cite{yang2014scalable,iluk2019miller}).
We solve a 2D stationary INS with $\text{Re}$ up to 5000 and a 3D
stationary INS with Re 20, with degrees of freedom up to about 2.4
million and 10 million, respectively. When applicable, we discretized
the INS equations using $P_{2}$-$P_{1}$ Taylor-Hood (TH) elements\cite{taylor1973numerical}
in all the comparisons, which are inf-sup stable.\cite{boffi2013mixed}
We used double-precision floating-point arithmetic with a nonlinear
residual of $10^{-5}$ or $10^{-6}$. Unless otherwise noted, we used
(F)GMRES with restart 30 and limited the maximum number of (F)GMRES
iteration to 200 within each nonlinear step. For HILUNG, we set $\epsilon$
to $0.8$ in line\ \ref{line:refactor} to trigger the factorization
of $\boldsymbol{J}_{S}$ when the solution changes rapidly, and we
set $\beta$ to $0.05$ to switch from Picard iterations to Newton
iterations in line~\ref{line:hot-start}, while still using the Oseen
operator as the sparsifier $\boldsymbol{J}_{S}$ when computing the
preconditioner for Newton iterations. We conducted our tests on a
single core of a cluster running CentOS 7.4 with dual 2.5 GHz Intel
Xeon CPU E5-2680v3 processors and 64 GB of RAM. All compute-intensive
kernels in HILUNG were implemented in C++, compiled by GCC 4.8.5 with
optimization flag `-O3', and then built into a MEX function for access
through MATLAB R2020a with MATLAB's built-in Intel MKL for LAPACK.
We used the same optimization flag and linked them with MKL either
with MATLAB's built-in version (for ILUPACK) with the standalone MKL
2018 (for MUMPS).

\subsection{2D driven-cavity problem\label{subsec:2D-drive-cavity-problem}}

We first assess HILUNG using the 2D lid-driven cavity problem over
the domain $\left[-1,1\right]^{2}$ using a range of Re and mesh resolutions.
This problem is widely used in the literature,\cite{ghia1982high,elman2006block,elman2014finite}
so it allows us to perform quantitative comparisons. The kinetic viscosity
is equal to $\nu=2/\text{Re}$. The no-slip wall condition is imposed
along all sides except for the top wall. There are two commonly used
configurations for the top wall: 1) The \emph{standard }top wall boundary
condition, which reads
\begin{equation}
\boldsymbol{u}_{\text{standard}}=\left[1,0\right],\label{eq:std-top-wall}
\end{equation}
and 2) a \emph{regularized }top wall boundary condition, such as\cite{elman2014finite}
\begin{equation}
\boldsymbol{u}_{\text{regularized}}=\left[1-x^{4},0\right].\label{eq:regularized-top-wall}
\end{equation}
 For this comparative study, we considered the standard boundary
condition \eqref{eq:std-top-wall}, which is more challenging to solve.
The pressure field has a ``do-nothing'' boundary condition, so the
coefficient matrix has a null space spanned by $\left[\boldsymbol{0},\boldsymbol{1}\right]^{T}$,
where the $\boldsymbol{1}$ components correspond to the constant
pressure mode (aka the hydrostatic pressure). We resolved the null
space in HILUNG as described in Section~\ref{subsec:Improving-robustness-with}.
Despite the simple geometry, the pressure contains two corner singularities
(cf. Figure~\ref{fig:pressure}), which become more severe as the
mesh is refined, leading to significant challenges for nonlinear solvers
and preconditioners. We used uniform meshes following the convention
as in Elman et al.,\cite{elman2014finite} except that we split the
$Q_{2}$ and $Q_{1}$ rectangular elements along one diagonal direction
to construct $P_{2}$ and $P_{1}$ triangular elements. We use level-$\ell$
mesh to denote the uniform mesh with $\left(2^{\ell-1}\right)^{2}$
$Q_{2}$ elements. For TH elements, there are $\left(2^{\ell}-1\right)^{2}$
DOFs in velocities and $\left(2^{\ell-1}+1\right)^{2}$ DOFs in pressure.
For the level-10 mesh, for example, there are approximately 2.36
million unknowns.

\subsubsection{Robustness of HILUNG}

\begin{figure}
\subfloat[\label{fig:driven-cavity-Ghia}Verification of velocities.]{\includegraphics[width=0.33\columnwidth]{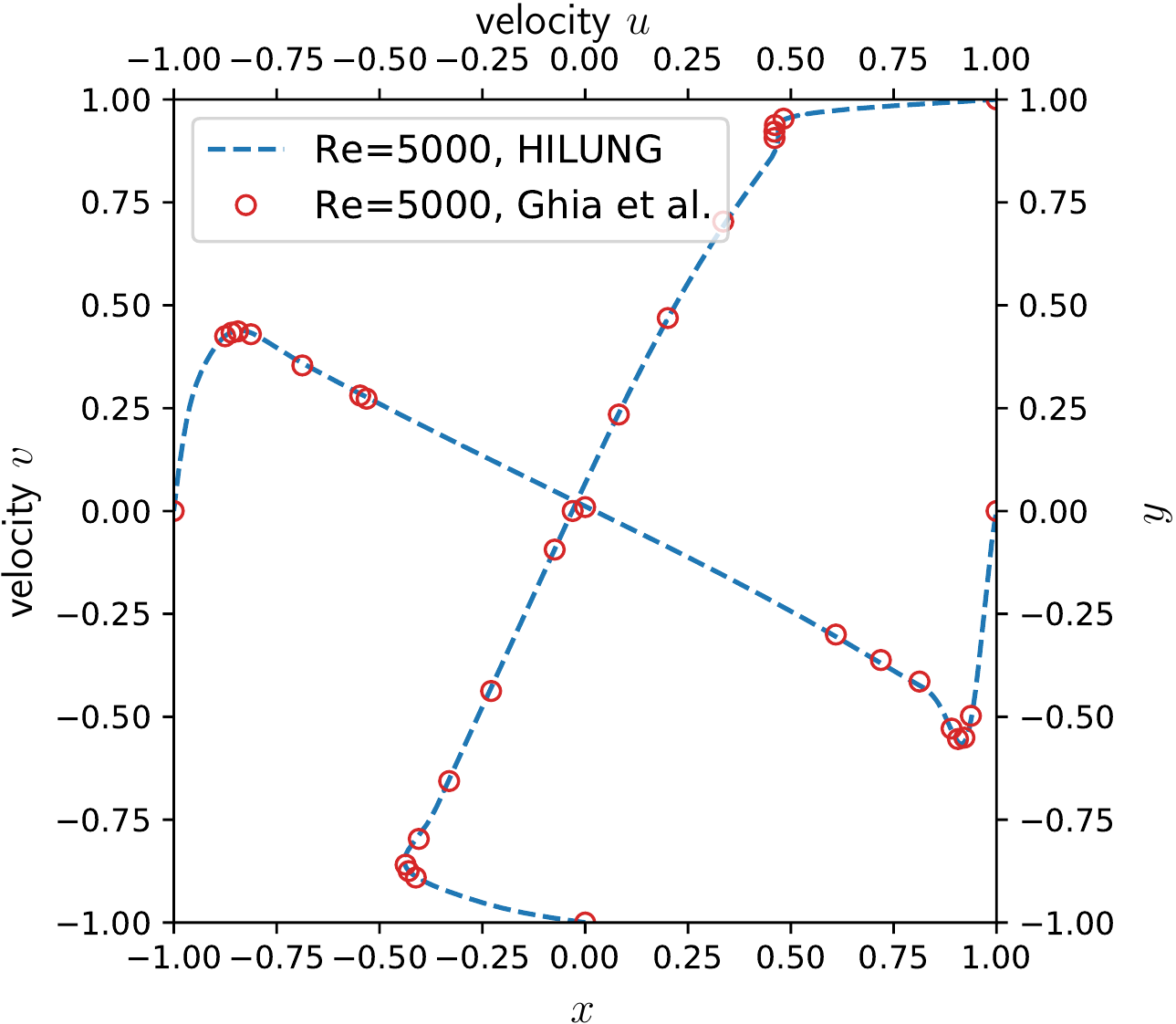}

}\hspace*{\fill}\subfloat[\label{fig:pressure}Solution of pressure.]{\includegraphics[width=0.33\columnwidth]{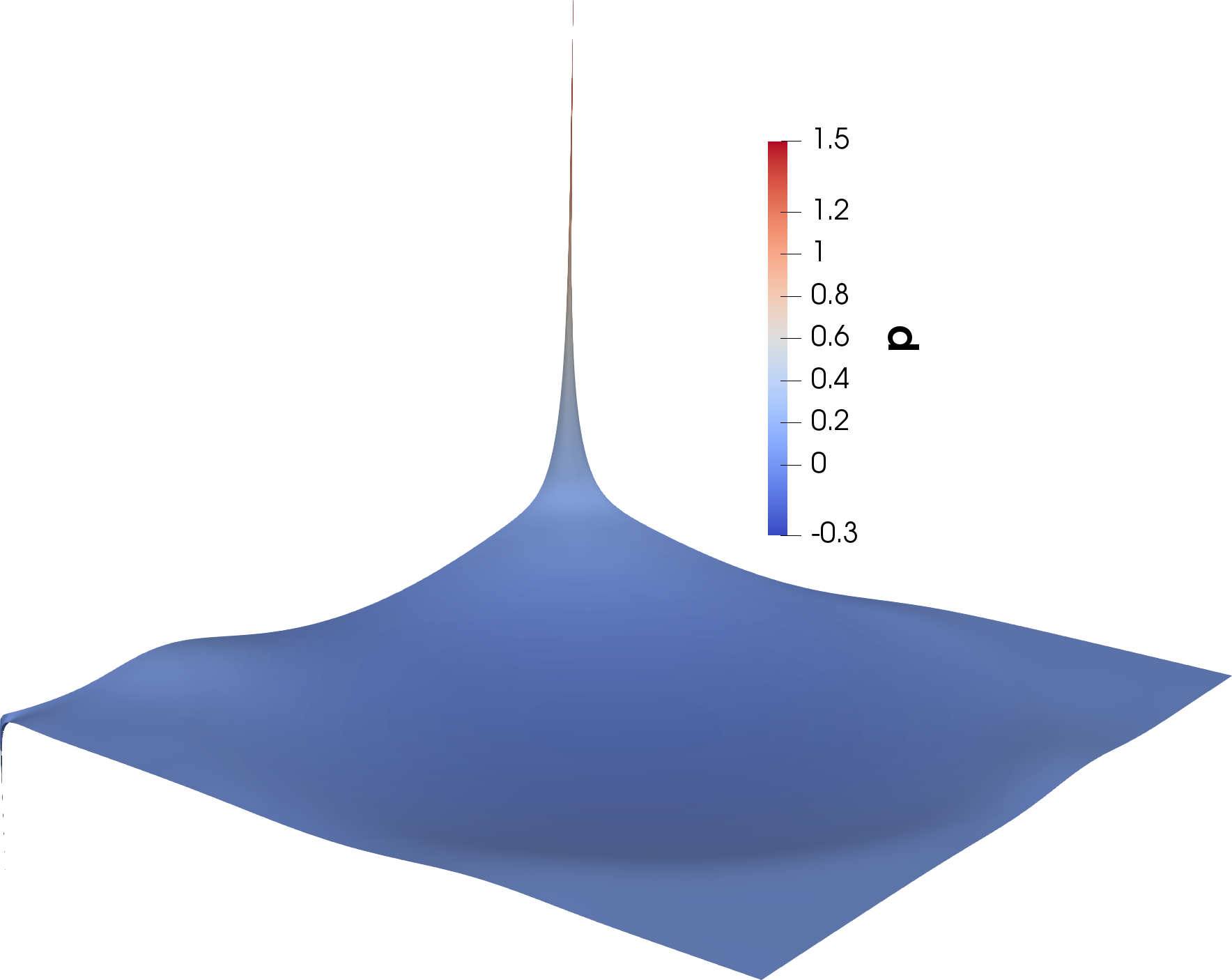}

}\hspace*{\fill}\subfloat[\label{fig:streamline}Demonstration of streamline.]{\includegraphics[width=0.33\columnwidth]{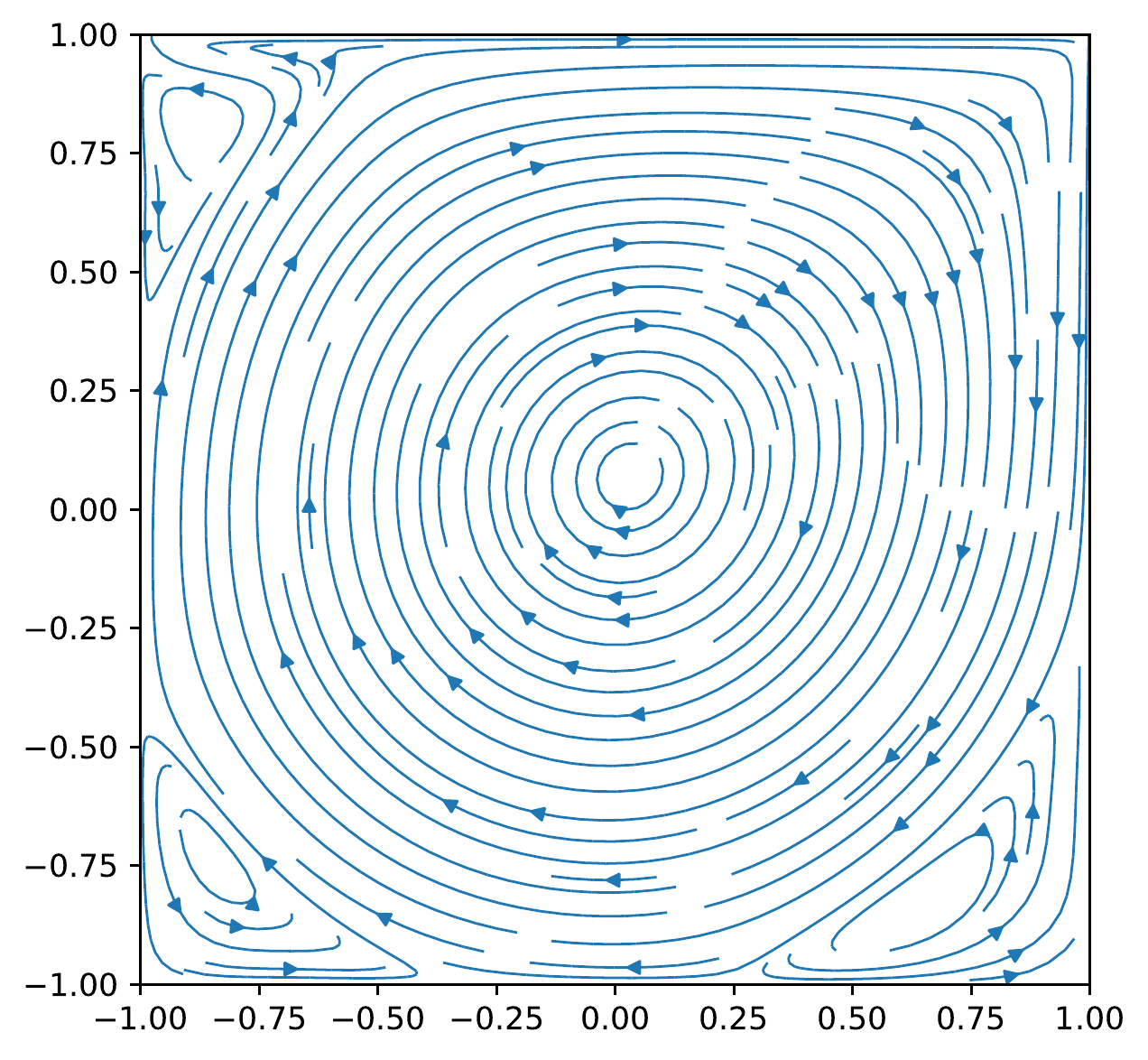}

}

\caption{2D driven-cavity problem with standard top-wall boundary condition
and Reynolds number $5000$. (a) Comparison of velocities along $x=0$
and $y=0$ with those by Ghia et al.\cite{ghia1982high} (b) Solutions
of pressure with two corner singularities. (c) The streamline plot,
which agrees very well to that of Ghia et al.\cite{ghia1982high}}
\end{figure}

\begin{figure}
\subfloat[\label{fig:Convergence-history-cavity2000}Convergence history of
2D driven-cavity with $\text{Re}=2000$.]{\includegraphics[width=0.48\columnwidth]{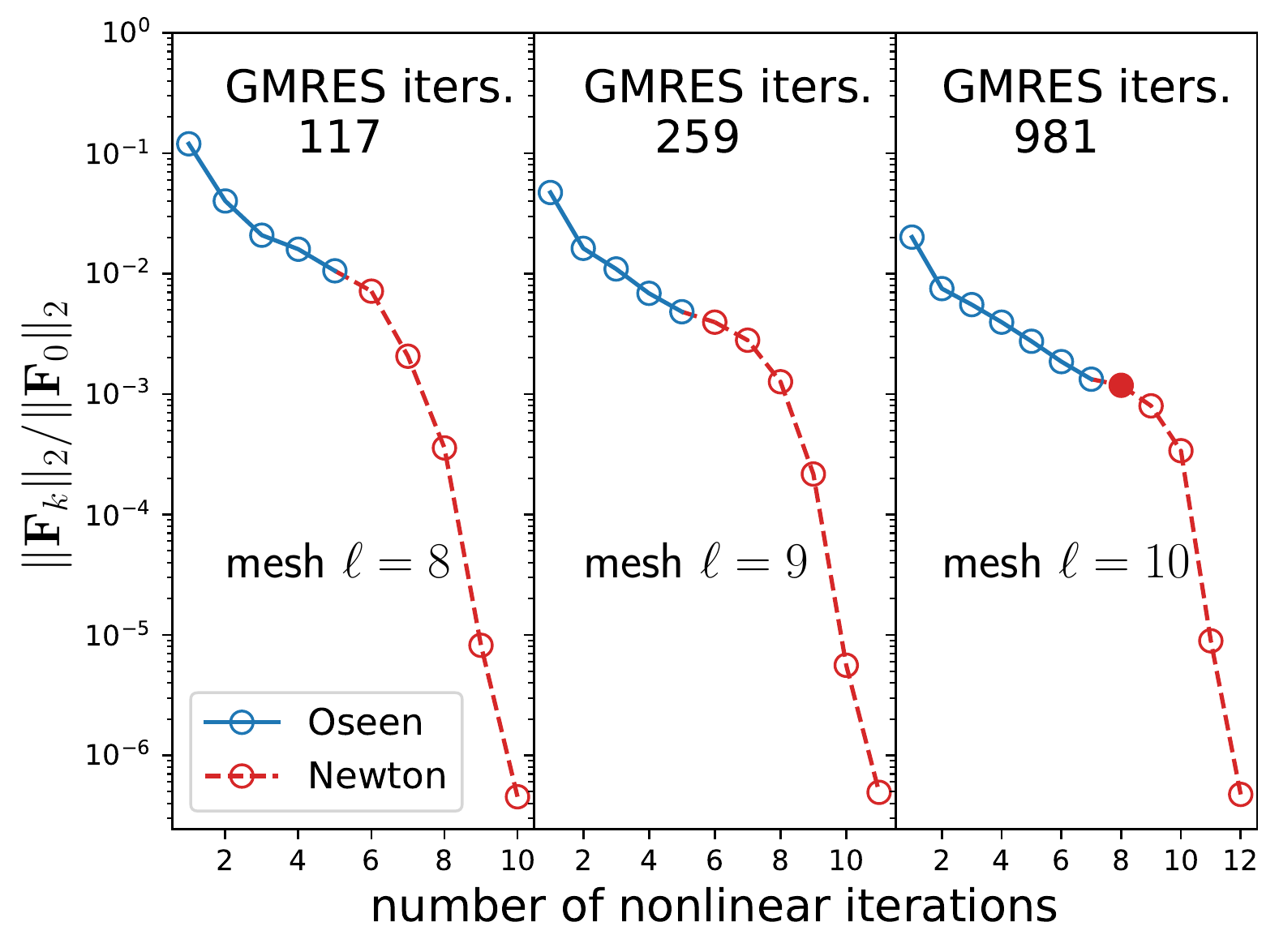}

}\hspace*{\fill}\subfloat[\label{fig:Convergence-history-cavity5000}Convergence history of
2D driven-cavity with $\text{Re}=5000$.]{\includegraphics[width=0.48\columnwidth]{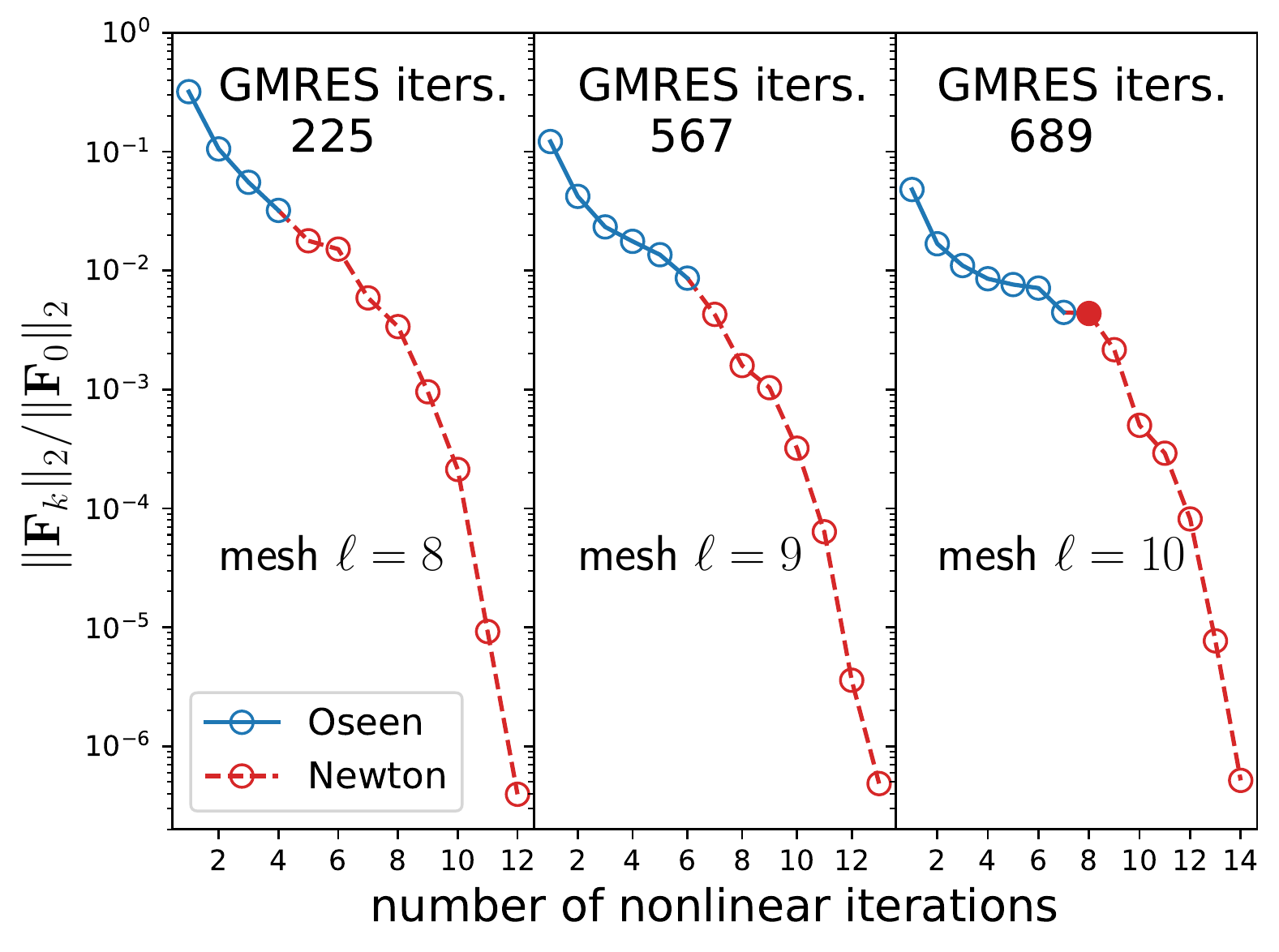}

}

\caption{\label{fig:Convergence-of-the}Convergence of the 2D driven-cavity
problems with different Reynolds numbers (a) 2000 and (b) 5000, where
solid dots (i.e., $\CIRCLE$) indicate that damping was invoked.}
\end{figure}

We first demonstrate the robustness of HILUNG for $\text{Re}=2000$
and $\text{Re}=5000$, which are moderately high and are challenging
due to the corner singularities in pressure (cf. Figure~\ref{fig:pressure}).
We chose nonlinear relative tolerance $\sigma=10^{-6}$ in \eqref{eq:term-nt},
and we used the solution of the Stokes equations as the initial guess
for nonlinear iterations in all cases. We set $N=20$ as the threshold
to trigger refactorization for the level-8 and 9 meshes, and we reduced
it to $N=15$ for the level-10 mesh due to the steeper corner singularities.
Figures\ \ref{fig:driven-cavity-Ghia} and~\ref{fig:streamline}
plot the velocities along the center lines $x=0$ and $y=0$ and the
streamline for $\text{Re}=5000$, which agreed very well with the
results of Ghia et al.\cite{ghia1982high} Figure\ \ref{fig:Convergence-of-the}
shows the convergence history of the nonlinear solvers on levels 8,
9, and 10 meshes, along with the total number of GMRES iterations.
The results indicate that HILUNG converged fairly smoothly under mesh
refinement. It is worth mentioning that no grad-div stabilization
was added in HILUNG to its coefficient matrix $\boldsymbol{J}$ or
the sparsifier $\boldsymbol{J}_{S}$.


\subsubsection{Effects of adaptive factorization and iterative refinement}

We then assess the effectiveness of adaptive refactorization (AR)
and iterative refinement (IR) in HILUNG. In our experiments, IR did
not improve Picard iterations, so we applied it only to Newton iterations.
When IR is enabled, it incurs an extra matrix-vector multiplication.
Hence, when IR is disabled, we doubled the upper limit of GMRES iterations
per nonlinear solver to 400 and doubled the parameter $N$ to $40$
for triggering refactorization. Table\ \ref{tab:HILUNG-effects}
compares the total runtimes and the numbers of GMRES iterations with
both AR and IR enabled, with only AR, and with only IR and with refactorization
at each each nonlinear iteration. It can be seen that AR was effective
in reducing the overall runtimes for all cases because the HILUCSI
factorization is more costly than triangular solves. Overall, enabling
both AR and IR delivered the best performance, especially on finer
meshes. IR was effective on the level-9 mesh. Compared to enabling
AR alone, enabling both IR and AR improved runtimes by about 10\%
for $\text{Re}=1000$ and $2000$ and about 30\% for $\text{Re}=5000$.
\begin{table}
\caption{\label{tab:HILUNG-effects}Comparison of total runtimes in seconds
and numbers of nonlinear iterations (in parentheses) with both adaptive
refactorization (AR) and iterative refinement (IR) (denoted as AR+IR)
vs. with only AR and only IR for 2D driven cavity problem with nonlinear
relative tolerance $10^{-6}$. Leaders are in boldface.}

\centering{}\setlength\tabcolsep{2pt}%
\begin{tabular}{ccccccccccccc}
\toprule 
\multirow{2}{*}{$\ell$} &  & \multicolumn{3}{c}{$\text{Re}=1000$} &  & \multicolumn{3}{c}{$\text{Re}=2000$} &  & \multicolumn{3}{c}{$\text{Re}=5000$}\tabularnewline
\cmidrule{3-5} \cmidrule{4-5} \cmidrule{5-5} \cmidrule{7-9} \cmidrule{8-9} \cmidrule{9-9} \cmidrule{11-13} \cmidrule{12-13} \cmidrule{13-13} 
 &  & AR+IR & AR & IR &  & AR+IR & AR & IR &  & AR+IR & AR & IR\tabularnewline
\midrule
7 &  & \textbf{10.2} (8) & 10.9 (8) & 16.5 (8) &  & \textbf{8.9} (9) & 11 (9) & 16 (9) &  & \textbf{25.8} (15) & 38.0 (16) & 55.6 (15)\tabularnewline
8 &  & 55.1 (10) & \textbf{54.2} (10) & 96.2 (10) &  & \textbf{46} (10) & 49 (10) & 91 (10) &  & \textbf{146} (12) & 213 (13) & 194 (12)\tabularnewline
9 &  & \textbf{489} (10) & 551 (10) & 517 (10) &  & \textbf{335} (11) & 397 (10) & 429 (11) &  & \textbf{635} (13) & 1.2e3(14) & 841 (13)\tabularnewline
10 &  & \textbf{5.2e3} (12) & 6.6e3 (16) & \textbf{5.2e3} (12) &  & \textbf{4.3e3} (12) & 6.1e3 (17) & \textbf{4.3e3} (12) &  & 4.2e3 (14) & 5.7e3 (16) & \textbf{4.1e3} (14)\tabularnewline
\bottomrule
\end{tabular}
\end{table}

\subsubsection{Comparison with customized preconditioners for INS}

To evaluate HILUNG with the state-of-the-art customized preconditioners
for INS, we compare it with four popular choices, including the augmented
Lagrangian (AL) preconditioner as proposed by Benzi and Olshanskii,\cite{benzi2006augmented}
the modified augmented Lagrangian (MAL) preconditioner as proposed
by Benzi, Olshanskii, and Wang,\cite{benzi2011modified} the \emph{pressure
convection diffusion (PCD)},\cite{silvester2001efficient,kay2002preconditioner}
and \emph{least-squares commutator} \emph{(LSC)}.\cite{elman2006block}
We chose these preconditioners for comparisons since they have all
been adopted in some recent open-source software packages,\cite{bootland2019preconditioners,farrel2019alcode,farrell2019augmented,ifiss}
so they represent the state of the art and we can use the existing
software directly for a fair comparison. In particular, for AL, we
used ALFI based on FireDrake by Farrell et al.,\cite{farrell2019augmented}\footnote{We used the predecessor of ALFI as described by Farrell et al.\cite{farrell2019augmented,farrel2019alcode}
since the most recent ALFI did not work with the software components
described therein and no proper name was given to its predecessor.} which uses a customized geometric multigrid method for the leading
block. For MAL, we used the implementation based on FreeFEM by Moulin
et al.,\cite{moulin2019augmented} hereafter referred to as MAL-FF,
which uses the direct solver MUMPS v5.0.2 with OpenBLAS for complete
factorization and solve of the leading block. For PCD and LSC, we
used the MATLAB implementation in IFISS v3.6,\cite{ers14,ifiss} which
uses complete factorization for the leading block. We used the same
uniform meshes for HILUNG, ALFI, MAL-FF, and IFISS. However, we used
$P_{2}$-$P_{1}$ elements for HILUNG and MAL-FF, used $P_{2}$-$P_{0}$
elements with ALFI,\cite{farrell2019augmented,farrel2019alcode} and
used $Q_{2}$-$Q_{1}$ TH elements with IFISS without subdividing
the quadrilaterals. This apparent discrepancy is because neither ALFI\cite{farrell2019augmented,farrel2019alcode}
nor IFISS supports $P_{2}$-$P_{1}$ TH elements. Although HILUNG
could converge to $10^{-6}$ for $\text{Re}=5000$ on the finest mesh,
as we have shown in Table~\ref{tab:HILUNG-effects}, we loosened
the nonlinear relative tolerance $\sigma$ from $10^{-6}$ to $10^{-5}$
for all the codes, since some other methods had difficulty converging
beyond $10^{-5}$ even for smaller Re. Whenever possible, we used
the default parameters in IFISS, ALFI, and MAL-FF. It is worth noting
that ALFI solved a sequence of problems with different Reynolds numbers
(e.g., $\text{Re}=[\text{\ensuremath{\epsilon}},10,100,200,\ldots,5000,\ldots]$
for $\epsilon\approx0$), and the solution obtained from a smaller
Re was used as the initial guess for the next one.\cite{farrell2019augmented}
However, such a ramping would require solving 50 equations before
solving for Re=5000, which is overly expensive computationally. Hence,
we excluded such a drastic ramping strategy and instead used the solution
of the Stokes problem as the initial guess for nonlinear iterations
for all the methods, as suggested by Elman et al.\cite{elman2014finite}
For ALFI, we set the penalty parameter to be $\gamma=10^{4}$, as
suggested by Farrell et al.\cite{farrell2019augmented} For MAL-FF,
we used $\gamma=0.3$, since we found that it performed better in
our tests than the ``optimal'' $\gamma\simeq1$ as suggested by
Moulin et al.\cite{moulin2019augmented} For HILUNG, ALFI, and IFISS,
we set the maximum number of (F)GMRES iterations per nonlinear iteration
to 200, while preserving its default value of 10000 in MAL-FF to avoid
premature termination. Note that IFISS only supports full GMRES (i.e.,
without restart); for the other codes, we use restarted FGMRES with
restart 30.

Table\ \ref{tab:Comparison-IFISS} compares the total numbers of
GMRES iterations and the numbers of nonlinear iterations for $\text{Re}=200$,
$1000$, and $5000$. IFISS failed on all the meshes for $\text{Re}=1000$
and $\text{Re}=5000$, so did MAL-FF for $\text{Re}=5000$, so we
omitted them from Table~\ref{tab:Comparison-IFISS}.\footnote{IFISS could solve the regularized driven-cavity problem with $\text{Re}=1000$
using the modified top-wall boundary condition \eqref{eq:regularized-top-wall}.\cite{elman2014finite}
Similarly, ALFI could solve a regularized version of the driven-cavity
problem for high $\text{Re}$.\cite{farrell2019augmented} } It is evident that HILUNG was the most robust in that it solved all
the problems. In contrast, ALFI and MAL-FF had comparable robustness,\footnote{ALFI could solve $\text{Re}=5000$ on the coarsest mesh in our tests
after we increased its maximum number of GMRES iterations per nonlinear
iteration from its default value of 100 to 200. Using ALFI's default
values, it solved the same number of problems as MAL-FF.} and PCD and LSC could not solve any of the problems with $\text{Re}\ge1000$.
In terms of the number of GMRES iterations, it can be seen that for
$\text{Re}=200$ ALFI and MAL-FF are less sensitive to mesh resolutions
than HILUNG. However, ALFI appeared to be more sensitive to mesh resolution
than HILUCSI for $\text{Re}=1000$, probably due to the loss of robustness
of the customized multigrid solver in ALFI for higher Reynolds number.
In contrast, MAL-FF remained insensitive to mesh resolution for $\text{Re}=1000$,
probably due to its use of complete factorization of the leading block,
whereas HILUCSI uses an incomplete factorization. Nevertheless, HILUNG
required fewer numbers of GMRES iterations than ALFI and MAL-FF for
$\text{Re}=1000$. More importantly, HILUNG was the only one to solve
the problem for $\text{Re}=5000$ on all the meshes. The numbers of
GMRES iterations decreased for HILUNG as the meshes refined for $\text{Re}=5000$
up to the level-8 mesh, because the omitted term $\boldsymbol{u}\cdot\boldsymbol{\nabla}\boldsymbol{u}_{k}$
in the sparsifier $\boldsymbol{J}_{S}$ (i.e., the Oseen operator)
becomes less and less important as the mesh is refined, so $\boldsymbol{J}_{S}$
becomes better approximations to $\boldsymbol{J}$ as the mesh is
refined for high $\text{Re}$.

\begin{table}
\caption{\label{tab:Comparison-IFISS}Comparison of the number of nonlinear
iterations and total number of GMRES iterations for HILUNG vs. ALFI,\cite{farrell2019augmented,farrel2019alcode}
MAL-FF,\cite{moulin2019augmented} and IFISS v3.6\cite{ifiss} with
PCD and LSC preconditioners with nonlinear relative tolerance $10^{-5}$.
Numbers in parentheses indicate the number of nonlinear iterations.
Leaders are in boldface. `$\times$' indicates that the nonlinear
solver failed to converge. MAL failed for $\text{Re}=5000$, whereas
PCD and LSD failed for $\text{Re}=1000$ and $5000$, so we omit their
corresponding columns.}

\centering{}%
\begin{tabular}{cccccccccccccc}
\toprule 
\multirow{2}{*}{$\ell$} &  & \multicolumn{5}{c}{$\text{Re}=200$} &  & \multicolumn{3}{c}{$\text{Re}=1000$} &  & \multicolumn{2}{c}{$\text{Re}=5000$}\tabularnewline
\cmidrule{3-7} \cmidrule{4-7} \cmidrule{5-7} \cmidrule{6-7} \cmidrule{7-7} \cmidrule{9-11} \cmidrule{10-11} \cmidrule{11-11} \cmidrule{13-14} \cmidrule{14-14} 
 &  & HILUNG & ALFI & MAL-FF & PCD & LSC &  & HILUNG & ALFI & MAL-FF &  & HILUNG & AL\tabularnewline
\midrule
6 &  & \textbf{8 }(5) & 23 (4) & 25 (4) & 159 (5) & 97 (5) &  & \textbf{18 }(7) & 44 (6) & 161 (8) &  & \textbf{325} (12) & 426 (25)\tabularnewline
7 &  & 23\textbf{ }(5) & \textbf{22} (4) & 25 (4) & 147 (5) & 161 (6) &  & \textbf{36 }(8) & 73 (8) & 110 (6) &  & \textbf{231} (15) & $\times$\tabularnewline
8 &  & 81 (7) & 20 (3) & \textbf{19} (3) & 154 (5) & 148 (5) &  & \textbf{98 }(10) & 757 (12) & 161 (8) &  & \textbf{197} (11) & $\times$\tabularnewline
\bottomrule
\end{tabular}
\end{table}

While the number of GMRES iterations is an important measure of the
overall efficiency, it is also important to compare the overall computational
cost. For completeness, we report detailed timing comparisons for
HILUNG, ALFI, and MAL-FF in Table~\ref{tab:Timing-comparison}. We
omit IFISS for timing comparison since it is implemented in MATLAB,
it uses full GMRES without restart, and its ``ideal setting'' uses
complete factorization. From Table~\ref{tab:Timing-comparison},
it is evident that HILUNG is the overall winner in terms of runtime,
especially for high Re. In particular, for $\text{Re}=1000$, HILUNG
was a factor of two to 40 faster than ALFI and a factor of 1.37--1.75
faster than MAL-FF. Compared to ALFI, the significantly better performance
of MAL-FF was probably due to the excellent cache performance of MUMPS
for 2D problems. For the same reason, MAL-FF was even faster than
HILUNG on the two finer meshes for $\text{Re}=200$. However, as we
will show in Section~\ref{subsec:3D-laminar-flow}, MUMPS performs
significantly worse for 3D problems due to its increased space and
time complexity. For Re=5000, since only HILUNG solved all the problems,
only as a point of reference Table~\ref{tab:Timing-comparison} reports
the runtimes for ALFI and MAL-FF before they reported failures, which
took a factor of 4.5 to 40 longer than the successful runs of HILUNG.

\begin{table}
\caption{\label{tab:Timing-comparison}Comparison of total runtimes in seconds
of HILUNG vs. ALFI\cite{farrell2019augmented,farrel2019alcode} and
MAL-FF.\cite{moulin2019augmented} The cases with `{*}' indicate the
runtimes before the code reported failure. Leaders are in boldface.}

\centering{}%
\begin{tabular}{ccccccccccccc}
\toprule 
\multirow{2}{*}{$\ell$} &  & \multicolumn{3}{c}{$\text{Re}=200$} &  & \multicolumn{3}{c}{$\text{Re}=1000$} &  & \multicolumn{3}{c}{$\text{Re}=5000$}\tabularnewline
\cmidrule{3-5} \cmidrule{4-5} \cmidrule{5-5} \cmidrule{7-9} \cmidrule{8-9} \cmidrule{9-9} \cmidrule{11-13} \cmidrule{12-13} \cmidrule{13-13} 
 &  & HILUNG & ALFI & MAL-FF &  & HILUNG & ALFI & MAL-FF &  & HILUNG & ALFI & MAL-FF\tabularnewline
\midrule
6 &  & \textbf{1.39} & 3.0 & 1.64 &  & \textbf{2.13} & 4.23 & 3.74 &  & \textbf{7.21} & 32.6 & 75.2{*}\tabularnewline
7 &  & 7.95 & 14.0\textbf{} & \textbf{6.9} &  & \textbf{8.42} & 37.8 & 11.6 &  & \textbf{25.6} & 813{*} & 281{*}\tabularnewline
8 &  & 40.8 & 59.4\textbf{} & \textbf{25.1} &  & \textbf{41.9} & 1.7e3 & 67.6 &  & \textbf{124} & 4.4e4{*} & 1.2e3{*}\tabularnewline
\bottomrule
\end{tabular}
\end{table}

In this 2D comparison, the superior robustness and efficiency of HILUNG
were primarily due to the use of HILUCSI as its preconditioner, which
enabled robust convergence of GMRES in the inner iteration. This conclusion
is supported by the study, since almost all the methods had a comparable
number of nonlinear iterations, and the main differences are the average
numbers of GMRES iterations. Furthermore, MAL-FF failed for $\text{Re}=5000$
due to the lack of convergence of GMRES in its inner iteration, even
when its maximum number of GMRES iterations was set to 10000, probably
due to its poor approximation to the Schur complement with a small
$\gamma$; similarly, IFISS also seems to suffer from an inaccurate
approximation to the Schur complement, even for $\text{Re}=1000$,
despite its use of complete factorization and full GMRES. For ALFI,
GMRES also converged slowly on finer meshes for higher Re, probably
because the large grad-div penalty term $\gamma\boldsymbol{\nabla}\boldsymbol{\nabla}\cdot\boldsymbol{u}$
with $\gamma=10^{4}$ in AL led to ill-conditioning of the overall
system, making it difficult for the pressure to converge, especially
near the corner singularities. HILUCSI does not suffer from any of
these deficiencies, although its scalability is not as good as a multigrid
method for low $\text{Re}$.

\subsection{\label{subsec:3D-laminar-flow}3D laminar flow over cylinder}

\begin{figure}
\subfloat[\label{fig:3d-cylinder-cfg}Computational domain.]{\includegraphics[width=0.41\columnwidth]{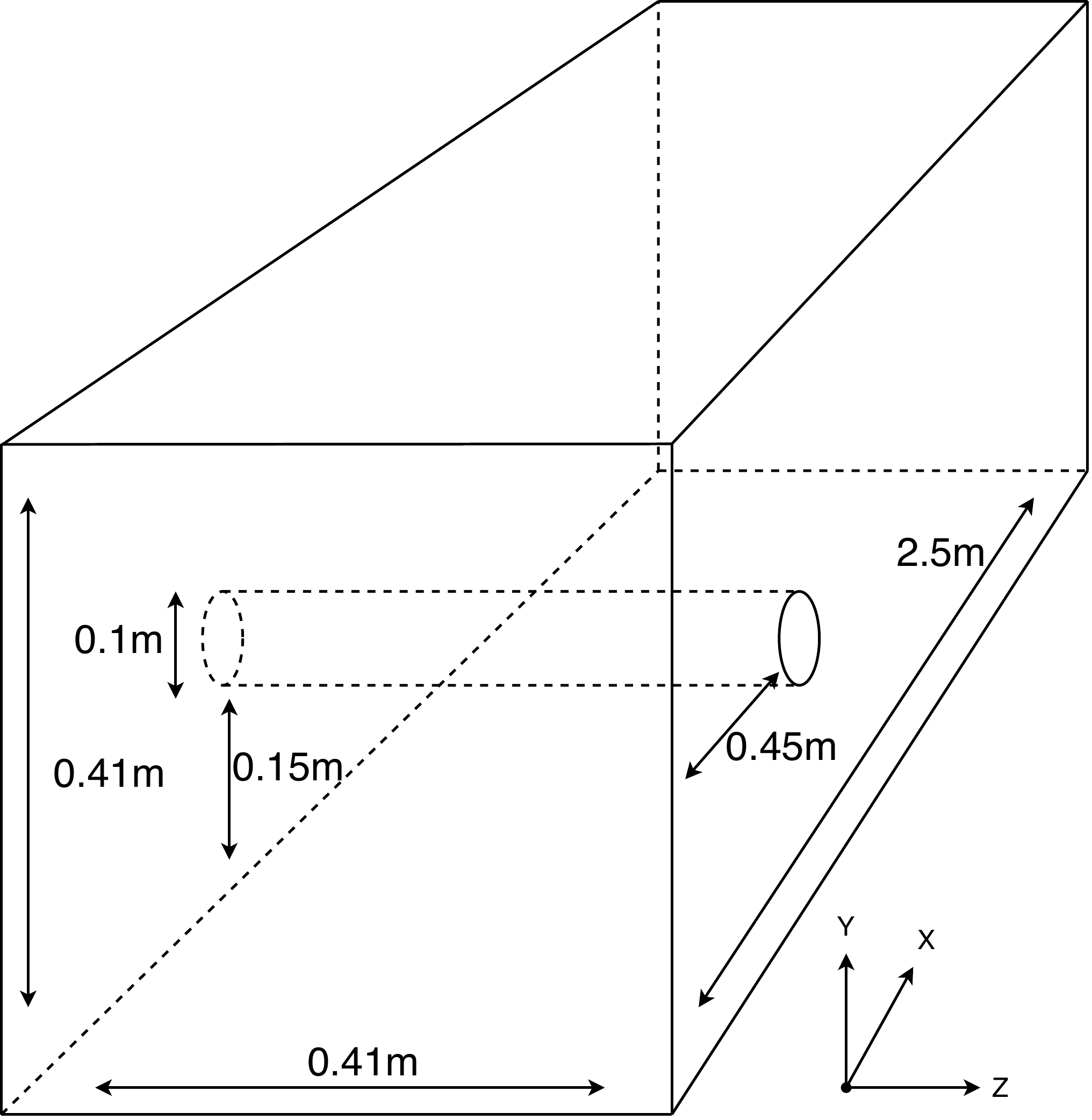}

}\hfill{}\subfloat[\label{fig:prob_samples}Example graded coarse mesh (top) and the
computed fluid speed (bottom).]{\includegraphics[width=0.48\columnwidth]{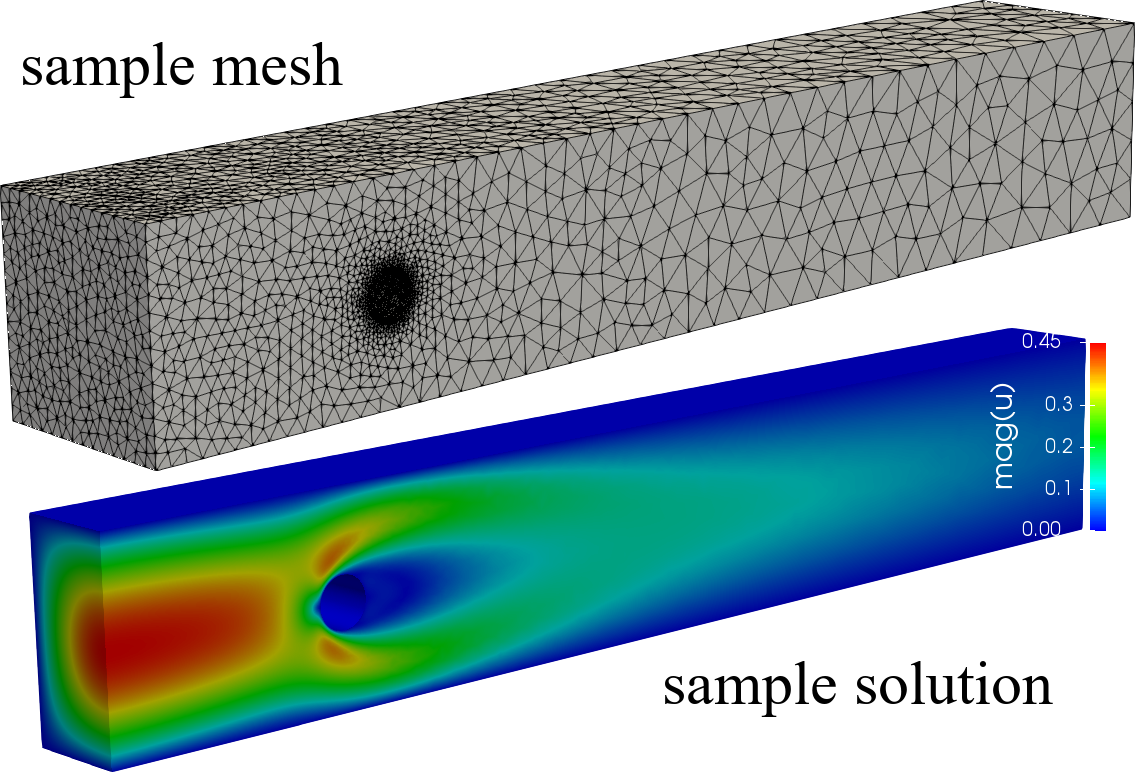}}\caption{Computational domain (a) and sample mesh and solution (b) of 3D flow-over-cylinder
problem. The front and back faces correspond to the inflow and outflow
walls, respectively.}
\end{figure}

Since real-world fluid problems are typically 3D, we also assess HILUNG
for a 3D problem, namely the flow-over-cylinder problem described
by Sch\"afer and Turek.\cite{schafer1996benchmark} The computation
domain is shown in Figure~\ref{fig:3d-cylinder-cfg}. The inflow
(front face) reads $\boldsymbol{u}=\left[U(y,z),0,0\right]^{T}$ with
$U(y,z)=16\times0.45\thinspace yz\left(H-y\right)\left(H-z\right)/H^{4}$,
where $H=0.41$ is the height and width of the channel. A ``do-nothing''
velocity is imposed for the outflow along with zero pressure. The
no-slip wall condition is imposed on the top, bottom, and cylinder
faces. The Reynolds number of this problem is given by $\text{Re}=4\times0.45\thinspace D/(9\thinspace\nu)=20$,
where $D=0.1$ and $\nu=1\times10^{-3}$ are the cylinder diameter
and kinetic viscosity, respectively. We generated four tetrahedral
meshes of different resolutions using Gmsh.\cite{geuzaine2009gmsh}
Despite small Re, the small viscosity leads to strongly asymmetric
and indefinite Jacobian matrices, making it difficult for Newton's
method to converge. Table\ \ref{tab:Problem-statistics} shows the
statistics of the matrices, where the largest system has about $10$
million DOFs and $907$ million nonzeros. Due to the global convergence
property of Picard iterations with the Oseen systems\cite[Chpt. 8]{elman2014finite}
and their use to hot start Newton's method in HILUNG, HILUNG converges
robustly for this problem on all the meshes. Figure\ \ref{fig:prob_samples}
shows a sample mesh and the fluid speed computed using HILUNG. 

\begin{table}
\caption{\label{tab:Problem-statistics}Statistics of different levels of meshes
for 3D flow, where nnz\_J and nnz\_S indicate number of nonzeros in
Newton and Oseen operators, respectively.}

\centering{}%
\begin{tabular}{cccccc}
\toprule 
 &  & mesh 1 & mesh 2 & mesh 3 & mesh 4\tabularnewline
\midrule
\#elems &  & 71,031 & 268,814 & 930,248 & 2,415,063\tabularnewline
\#unkowns &  & 262,912 & 1,086,263 & 3,738,327 & 9,759,495\tabularnewline
nnz\_J &  & 21,870,739 & 98,205,997 & 343,357,455 & 906,853,456\tabularnewline
nnz\_S &  & 9,902,533 & 43,686,979 & 152,438,721 & 401,879,584\tabularnewline
\bottomrule
\end{tabular}
\end{table}

\subsubsection{Efficiency comparison with other general-purpose approximate factorizations}

For the comparative study of the 2D problem in Section~\ref{subsec:2D-drive-cavity-problem},
the use of HILUCSI was the major factor for the robustness and efficiency
of HILUNG. Since HILUCSI is a general-purpose approximate factorization,
we focus on comparing HILUCSI with other general-purpose approximate
factorizations for this 3D comparative study, including ILUPACK v2.4,\cite{bollhofer2006ilupack}
and single-precision MUMPS v5.3.5.\cite{amestoy2000mumps,amestoy2019performance}
ILUPACK is most related to HILUCSI in that it also uses multilevel
ILU. In both HILUCSI and ILUPACK, we used $\text{droptol}$ $0.02$
and $0.01$ during Picard and Newton iterations, respectively, and
used their respective default options for the other parameters. To
be comparable with single-precision MUMPS, we also enabled mixed-precision
computations in both HILUCSI and ILUPACK. For completeness, we also
included ILU($1$) and ILU($2$) in our comparison due to its popularity.\cite{yang2014scalable,iluk2019miller}
Since none of these preconditioners has a corresponding nonlinear
solver, for a fair comparison, we used the same nonlinear solver as
described in Algorithm~\ref{alg:hilung} and simply replaced HILUCSI
with these approximate-factorization techniques as the preconditioner
for GMRES. In particular, we used GMRES(30) in the inner iteration
and used relative tolerance of $10^{-6}$ for the nonlinear iteration.

Table~\ref{tab:Comparison-3d-approx-factor} reports the factorization
times, total runtimes, and the numbers of nonlinear and GMRES iterations
for the four meshes. Table~\ref{tab:Comparison-3d-approx-factor}
shows that nonlinear solver failed to converge for any of the meshes
for GMRES with ILU($1$) and ILU(2), even after applying equilibration\cite{duff2001algorithms}
and fill-reduction reordering\cite{amestoy1996approximate} to improve
its robustness, despite the global convergence property of Picard
iterations with the Oseen systems and hot-started Newton iterations.
For the two coarser meshes, compared to ILUPACK, HILUCSI was about
a factor of 10 and 34 faster in terms of factorization cost and a
factor of 3.4 and 6.6 faster overall. MUMPS under-performed HILUCSI
in terms of both factorization and overall times, although it reduced
the number of GMRES iterations. In addition, we also tested the multi-threaded
MUMPS on 24 cores, which also under-performed the serial HILUCSI by
a factor of 1.4--1.7 in the overall runtimes for the two coarser
meshes due to the poor speedup of triangular solves, which dominate
the overall runtimes.

\begin{table}
\caption{\label{tab:Comparison-3d-approx-factor}Comparison of factorization
times, total times, and the number of GMRES and nonlinear iterations
for approximate-factorization-based preconditioners for the 3D flow-over-cylinder
problem. `$\times$' indicates nonconvergence of nonlinear iterations
with ILU(1) and ILU(2), and `$-$' indicates that factorization ran
out of the 64GB main memory. Numbers in the parentheses indicate the
number of nonlinear iterations. Leaders are in boldface.}

\centering{}\setlength\tabcolsep{2pt}%
\begin{tabular}{cccccccccccccccc}
\toprule 
\multirow{2}{*}{prec.} &  & \multicolumn{4}{c}{factorization times (s)} &  & \multicolumn{4}{c}{total times (s)} &  & \multicolumn{4}{c}{GMRES and Newton iters.}\tabularnewline
\cmidrule{3-6} \cmidrule{4-6} \cmidrule{5-6} \cmidrule{6-6} \cmidrule{8-11} \cmidrule{9-11} \cmidrule{10-11} \cmidrule{11-11} \cmidrule{13-16} \cmidrule{14-16} \cmidrule{15-16} \cmidrule{16-16} 
 &  & mesh 1 & mesh 2 & mesh 3 & mesh 4 &  & mesh 1 & mesh 2 & mesh 3 & mesh 4 &  & mesh 1 & mesh 2 & mesh 3 & mesh 4\tabularnewline
\midrule
HILUCSI &  & \textbf{17.5} & \textbf{78.3} & \textbf{255} & \textbf{659} &  & \textbf{71.5} & \textbf{485} & \textbf{1.8e3} & \textbf{4.2e3} &  & 135 (10) & 200 (13) & 199 (11) & 249 (11)\tabularnewline
ILU(1) or (2) &  & $\times$ & $\times$ & $\times$ & $\times$ &  & $\times$ & $\times$ & $\times$ & $\times$ &  & $\times$ & $\times$ & $\times$ & $\times$\tabularnewline
ILUPACK &  & 182 & 2.7e3 & $-$ & $-$ &  & 246 & 3.2e3 & $-$ & $-$ &  & 128 (9) & 150 (11) & $-$ & $-$\tabularnewline
MUMPS &  & 39.3 & 649 & $-$ & $-$ &  & 95.8 & 1.5e3 & $-$ & $-$ &  & \textbf{89} (10) & \textbf{127} (11) & $-$ & $-$\tabularnewline
\bottomrule
\end{tabular}
\end{table}

\subsubsection{Comparison of space and time complexities}

\begin{figure}
\subfloat[\label{fig:space-complexity}Space complexity.]{\includegraphics[width=0.48\columnwidth]{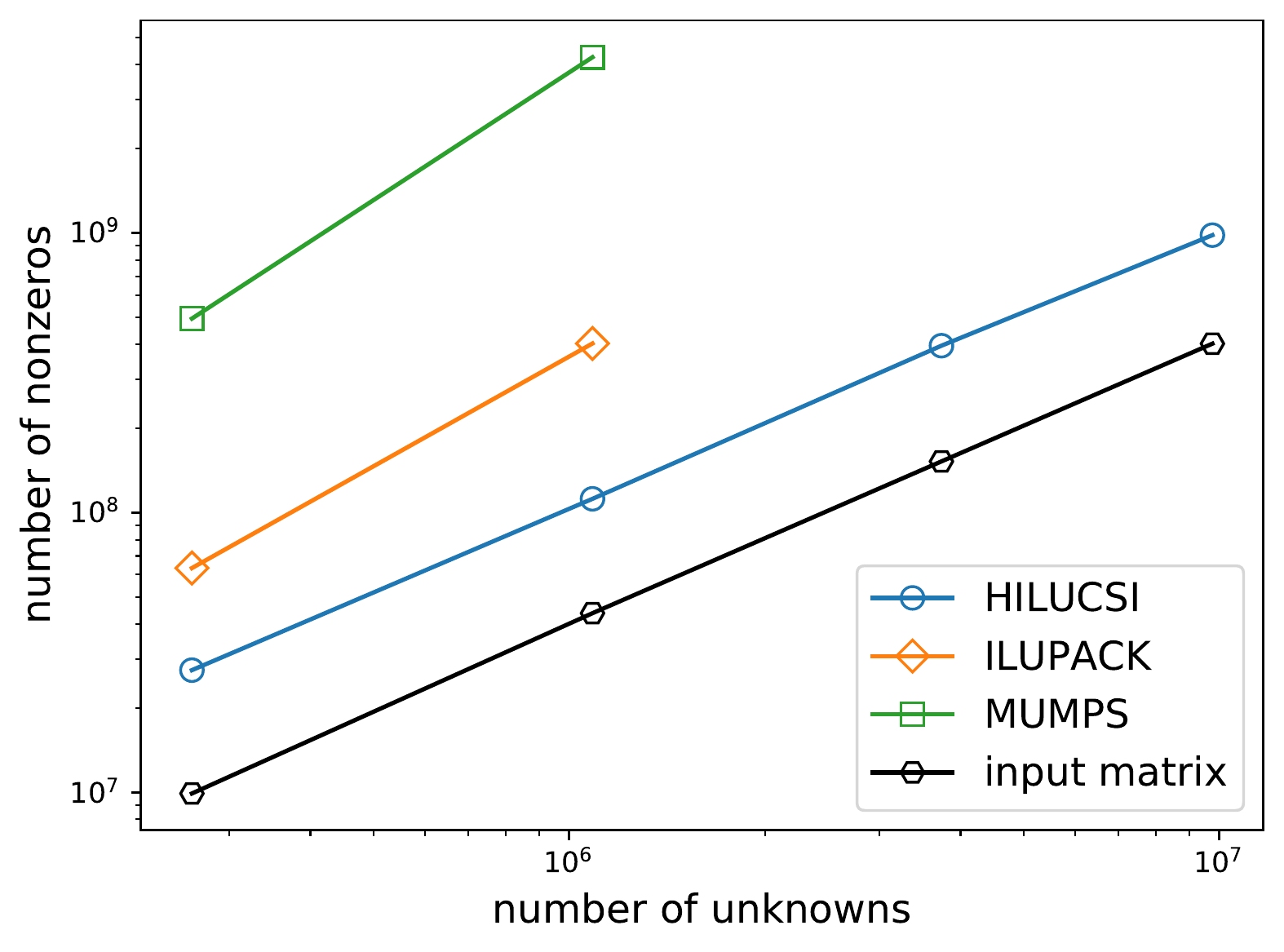}}\hfill{}\subfloat[\label{fig:time-complexity}Time complexity.]{\includegraphics[width=0.48\columnwidth]{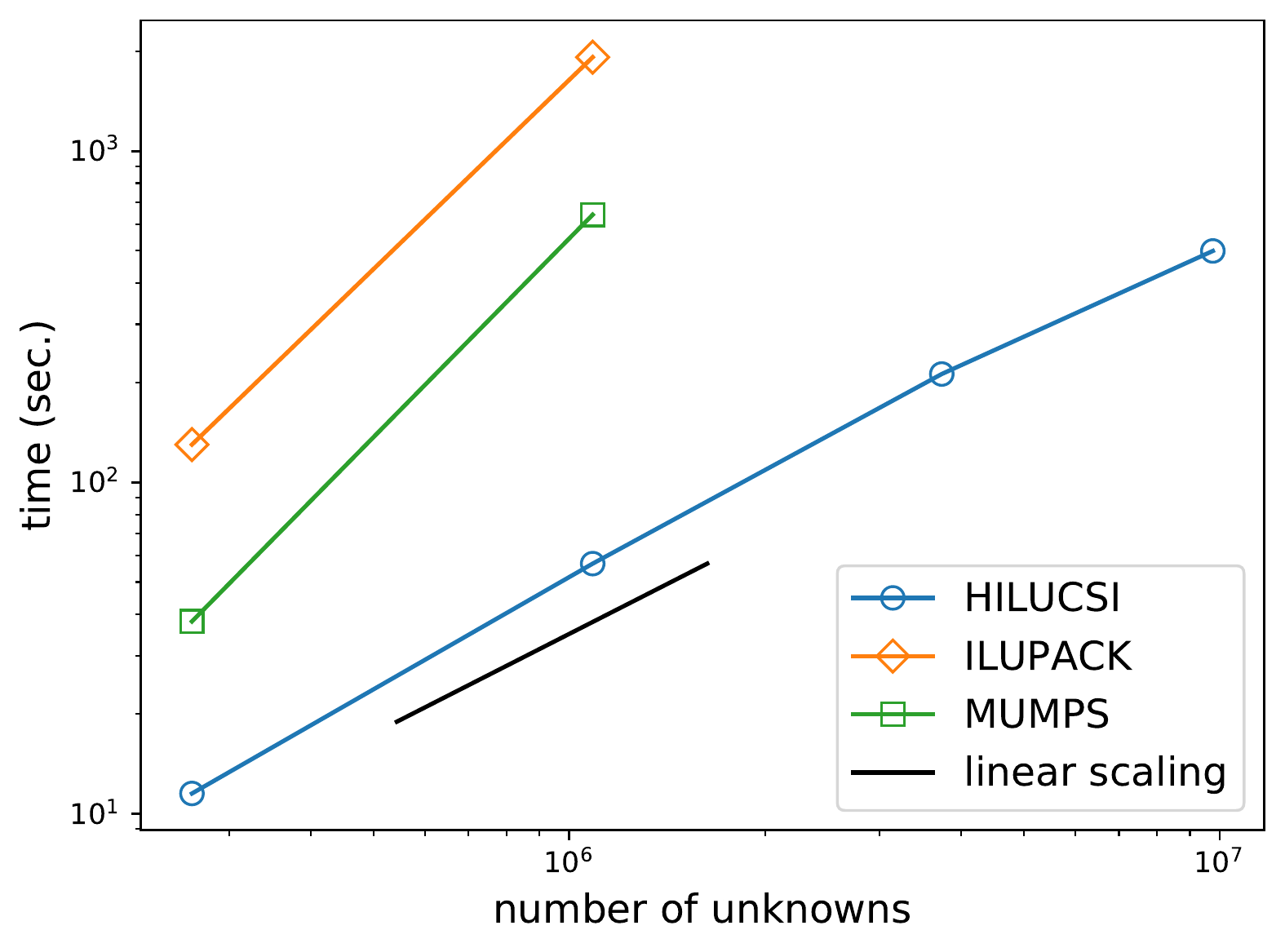}

}\caption{\label{fig:space-and-time-complexity}Comparison of (a) space and
(b) time complexities of approximate factorizations.}
\end{figure}

It is worth noting that both ILUPACK and single-precision MUMPS ran
out of the main 64 GB memory for the two larger meshes despite the
use of a sparsifier. If HILUCSI were applied to the full Jacobian
matrix for Newton's iterations, HILUNG would have also run out of
memory for the largest system. However, a more fundamental reason
for ILUPACK and MUMPS to run out of memory was that they both appear
to have superlinear space complexities, in that the number of nonzeros
in the (approximate) triangular factors grow super-linearly for ILUPACK
and MUMPS, as evident in Figure~\ref{fig:space-complexity}, which
shows the growth of nonzeros in the factorization of the Oseen operator
for HILUCSI, ILUPACK, and MUMPS, with $\text{droptol}=0.02$ for both
HILUCSI and ILUPACK. The superlinear space complexity also implies
that the factorization costs of ILUPACK and MUMPS are superlinear,
as evident in Figure~\ref{fig:time-complexity}. Such a superlinear
complexity limits the scalability of ILUPACK and MUMPS to large-scale
problems, even for the parallel version of MUMPS, regardless of whether
sparsifiers are used. In contrast, HILUCSI scaled linearly in both
memory and computational cost for its factorization and solve steps,
thanks to its scalability-oriented dropping in its multilevel structure.

\section{Conclusions}

\label{sec:Conclusions} In this work, we introduced \emph{HILUNG},
which is the first to incorporate a robust multilevel ILU preconditioner
into Newton-GMRES to solve the nonlinear equations from incompressible
Navier-Stokes equations. In particular, HILUNG applies HILUCSI on
a physics-aware sparsifying operator to compute a multilevel ILU.
Thanks to the scalability-oriented and inverse-based dual thresholding
in HILUCSI, HILUNG enjoys robust and rapid convergence of restarted
GMRES in its inner loops. Furthermore, by utilizing adaptive refactorization
and iterative refinement to improve runtime efficiency and leverages
some standard nonlinear techniques (including hot starting and damping),
HILUNG achieved high robustness and efficiency without suffering from
potential over-factorization. We demonstrated the effectiveness of
HILUNG for stationary incompressible Navier-Stokes equations using
Taylor-Hood elements. In addition, we compared HILUNG with some state-of-the-art
customized preconditioners, including two variants of augmented Lagrangian
preconditioners and two physics-based preconditioners, for the 2D
driven-cavity problem with the standard top-wall boundary condition.
We showed that HILUNG was the most robust in solving the problem with
Re $5000$ successfully on all levels of meshes and was also the most
efficient for Re $1000$. In addition, we demonstrated HILUNG in solving
the 3D flow-over-cylinder problem with ten million DOFs using only
60GB of RAM, while other alternatives failed to solve the problem
on the same system with less than four million DOFs. We also showed
that HILUNG outperformed another state-of-the-art multilevel ILU preconditioner
by more than an order of magnitude for a system with one million DOFs,
and even outperformed a state-of-the-art parallel direct solver on
24 cores for the same system. We have released the source code of
HILUCSI as part of the HIFIR project at \url{https://github.com/hifirworks/hifir}.
One limitation of this work is that HILUCSI is only serial. Future
research directions include parallelizing HILUCSI, applying it to
solve even higher-Re and larger-scale problems, and developing a custom
preconditioner for time-dependent INS with fully implicit Runge-Kutta
schemes.

\section*{Acknowledgments}

Computational results were obtained using the Seawulf computer systems
at the Institute for Advanced Computational Science of Stony Brook
University, which were partially funded by the Empire State Development
grant NYS \#28451. We thank the anonymous referees for their helpful
comments regarding the comparison of HILUNG with the state of the
art.

\section*{Data availability statement}

The core component of HILUNG, HILUCSI, is available as part of an
open-source library HIFIR at \url{https://github.com/hifirworks/hifir}.
Additional data that support the findings of this study are available
from the corresponding author upon reasonable request. 

\bibliography{refs}

\begin{thebibliography}{10}
\providecommand \doibase [0]{http://dx.doi.org/}%

\bibitem{turek1996comparative}
Turek S. A comparative study of time-stepping techniques for the incompressible
  {N}avier--{S}tokes equations: from fully implicit non-linear schemes to
  semi-implicit projection methods. {\it Int. J. Numer. Methods Fluids}
  1996\string; 22(10)\string: 987--1011.

\bibitem{elman2014finite}
Elman HC, Silvester DJ, Wathen AJ. {\it Finite {E}lements and {F}ast
  {I}terative {S}olvers: {W}ith {A}pplications {I}n {I}ncompressible {F}luid
  {D}ynamics}.
\newblock Oxford University Press, USA.
\newblock 2nd~ed. 2014.

\bibitem{bootland2019preconditioners}
Bootland N, Bentley A, Kees C, Wathen A. Preconditioners for two-phase
  incompressible {N}avier--{S}tokes flow. {\it SIAM J. Sci. Comput.}
  2019\string; 41(4)\string: B843--B869.

\bibitem{pearson2020preconditioners}
Pearson JW, Pestana J. Preconditioners for {K}rylov subspace methods: An
  overview. {\it GAMM-Mitteilungen} 2020\string; 43(4)\string: e202000015.

\bibitem{pernice2001multigrid}
Pernice M, Tocci MD. A multigrid-preconditioned {N}ewton--{K}rylov method for
  the incompressible {N}avier--{S}tokes equations. {\it SIAM J. Sci. Comput.}
  2001\string; 23(2)\string: 398--418.

\bibitem{heath2018scientific}
Heath MT. {\it Scientific {C}omputing: {A}n {I}ntroductory {S}urvey}. 80.
\newblock SIAM .
\newblock 2018.

\bibitem{brown1994convergence}
Brown PN, Saad Y. Convergence theory of nonlinear {N}ewton--{K}rylov
  algorithms. {\it SIAM J. Optim.} 1994\string; 4(2)\string: 297--330.

\bibitem{taylor1973numerical}
Taylor C, Hood P. A numerical solution of the {N}avier--{S}tokes equations
  using the finite element technique. {\it Comput. Fluids} 1973\string;
  1(1)\string: 73--100.

\bibitem{kelley1995iterative}
Kelley CT. {\it Iterative {M}ethods for {L}inear and {N}onlinear {E}quations}.
  16.
\newblock SIAM .
\newblock 1995.

\bibitem{benzi2005numerical}
Benzi M, Golub GH, Liesen J. Numerical solution of saddle point problems. {\it
  Acta Numerica} 2005\string; 14\string: 1--137.

\bibitem{brown1990hybrid}
Brown PN, Saad Y. Hybrid {K}rylov methods for nonlinear systems of equations.
  {\it SIAM J. Sci. Comput.} 1990\string; 11(3)\string: 450--481.

\bibitem{knoll2004jacobian}
Knoll DA, Keyes DE. Jacobian-free {N}ewton--{K}rylov methods: a survey of
  approaches and applications. {\it J. Comput. Phys.} 2004\string;
  193(2)\string: 357--397.

\bibitem{qin2000matrix}
Qin N, Ludlow DK, Shaw ST. A matrix-free preconditioned {N}ewton/{GMRES} method
  for unsteady {N}avier--{S}tokes solutions. {\it Int. J. Numer. Methods
  Fluids} 2000\string; 33(2)\string: 223--248.

\bibitem{saad2003iterative}
Saad Y. {\it Iterative {M}ethods for {S}parse {L}inear {S}ystems}. 82.
\newblock SIAM.
\newblock 2nd~ed. 2003.

\bibitem{pernice1998nitsol}
Pernice M, Walker HF. {NITSOL}: {A} {N}ewton iterative solver for nonlinear
  systems. {\it SIAM J. Sci. Comput.} 1998\string; 19(1)\string: 302--318.

\bibitem{gaston2009moose}
Gaston D, Newman C, Hansen G, Lebrun-Grandie D. {MOOSE}: {A} parallel
  computational framework for coupled systems of nonlinear equations. {\it
  Nucl. Eng. Des.} 2009\string; 239(10)\string: 1768--1778.

\bibitem{brune2015composing}
Brune PR, Knepley MG, Smith BF, Tu X. Composing scalable nonlinear algebraic
  solvers. {\it SIAM Rev.} 2015\string; 57(4)\string: 535--565.

\bibitem{balay2019petsc}
Balay S, Abhyankar S, Adams M, et al. {\it {PETS}c Users Manual}. Argonne
  National Laboratory; Argonne, IL:  2019.

\bibitem{ghia1982high}
Ghia U, Ghia KN, Shin C. High-{R}e solutions for incompressible flow using the
  {N}avier--{S}tokes equations and a multigrid method. {\it J. Comput. Phys.}
  1982\string; 48(3)\string: 387--411.

\bibitem{tuminaro2002backtracking}
Tuminaro RS, Walker HF, Shadid JN. On backtracking failure in {N}ewton--{G}MRES
  methods with a demonstration for the {N}avier--{S}tokes equations. {\it J.
  Comput. Phy.} 2002\string; 180(2)\string: 549--558.

\bibitem{persson2008newton}
Persson PO, Peraire J. Newton--{GMRES} preconditioning for discontinuous
  {G}alerkin discretizations of the {N}avier--{S}tokes equations. {\it SIAM J.
  Sci. Comput.} 2008\string; 30(6)\string: 2709--2733.

\bibitem{ur2008comparison}
{ur Rehman} M, Vuik C, Segal G. A comparison of preconditioners for
  incompressible {N}avier--{S}tokes solvers. {\it Int. J. Numer. Methods
  Fluids} 2008\string; 57(12)\string: 1731--1751.

\bibitem{benzi2006augmented}
Benzi M, Olshanskii MA. An augmented {L}agrangian-based approach to the {O}seen
  problem. {\it SIAM J. Sci. Comput.} 2006\string; 28(6)\string: 2095--2113.

\bibitem{benzi2011modified}
Benzi M, Olshanskii MA, Wang Z. Modified augmented {L}agrangian preconditioners
  for the incompressible {N}avier--{S}tokes equations. {\it Int. J. Numer.
  Methods Fluids} 2011\string; 66(4)\string: 486--508.

\bibitem{farrell2019augmented}
Farrell PE, Mitchell L, Wechsung F. An Augmented {L}agrangian Preconditioner
  for the 3D Stationary Incompressible {N}avier--{S}tokes Equations at High
  Reynolds Number. {\it SIAM J. Sci. Comput.} 2019\string; 41(5)\string:
  A3073--A3096.

\bibitem{moulin2019augmented}
Moulin J, Jolivet P, Marquet O. Augmented {L}agrangian preconditioner for
  large-scale hydrodynamic stability analysis. {\it Comput. Methods Appl. Mech.
  Eng.} 2019\string; 351\string: 718--743.

\bibitem{elman2006block}
Elman H, Howle VE, Shadid J, Shuttleworth R, Tuminaro R. Block preconditioners
  based on approximate commutators. {\it SIAM J. Sci. Comput.} 2006\string;
  27(5)\string: 1651--1668.

\bibitem{farrel2019alcode}
Software used in "{A}n augmented {L}agrangian preconditioner for the {3D}
  stationary incompressible {N}avier--{S}tokes equations at high {R}eynolds
  number". \url{https://doi.org/10.5281/zenodo.3247427};  2019

\bibitem{lee2015direct}
Lee M, Moser RD. Direct numerical simulation of turbulent channel flow up to
  {$Re_{\tau}\approx 5200$}. {\it J. Fluid Mech.} 2015\string; 774\string:
  395--415.

\bibitem{chen2021hilucsi}
Chen Q, Ghai A, Jiao X. {HILUCSI}: Simple, robust, and fast multilevel {ILU}
  for large-scale saddle-point problems from {PDE}s. {\it Numer. Linear Algebra
  Appl.} 2021.
\newblock \href {\doibase 10.1002/nla.2400} {doi: 10.1002/nla.2400}

\bibitem{ers14}
Elman H, Ramage A, Silvester D. {IFISS}: A computational laboratory for
  investigating incompressible flow problems. {\it SIAM Rev.} 2014\string;
  56\string: 261--273.

\bibitem{ifiss}
Silvester D, Elman H, Ramage A. {I}ncompressible {F}low and {I}terative
  {S}olver {S}oftware ({IFISS}) version 3.5.
  \url{http://www.manchester.ac.uk/ifiss};  2016.

\bibitem{bollhofer2011ilupack}
Bollh{\"o}fer M, Aliaga JI, Mart{\'\i}n AF, Quintana-Ort{\'\i} ES. {ILUPACK}.
  {\it Encyclopedia of Parallel Computing} 2011\string: 917--926.

\bibitem{dembo1982inexact}
Dembo RS, Eisenstat SC, Steihaug T. Inexact {N}ewton methods. {\it SIAM J.
  Numer. Anal.} 1982\string; 19(2)\string: 400--408.

\bibitem{eisenstat1994globally}
Eisenstat SC, Walker HF. Globally convergent inexact {N}ewton methods. {\it
  SIAM J. Optim.} 1994\string; 4(2)\string: 393--422.

\bibitem{eisenstat1996choosing}
Eisenstat SC, Walker HF. Choosing the forcing terms in an inexact {N}ewton
  method. {\it SIAM J. Sci. Comput.} 1996\string; 17(1)\string: 16--32.

\bibitem{dennis1996numerical}
Dennis~Jr JE, Schnabel RB. {\it Numerical {M}ethods for {U}nconstrained
  {O}ptimization and {N}onlinear {E}quations}. 16.
\newblock SIAM .
\newblock 1996.

\bibitem{an2007globally}
An HB, Bai ZZ. A globally convergent {N}ewton--{GMRES} method for large sparse
  systems of nonlinear equations. {\it Appl. Numer. Math} 2007\string;
  57(3)\string: 235--252.

\bibitem{bellavia2001globally}
Bellavia S, Morini B. A globally convergent {N}ewton-{GMRES} subspace method
  for systems of nonlinear equations. {\it SIAM J. Sci. Comput.} 2001\string;
  23(3)\string: 940--960.

\bibitem{murphy2000note}
Murphy MF, Golub GH, Wathen AJ. A note on preconditioning for indefinite linear
  systems. {\it SIAM J. Sci. Comput.} 2000\string; 21(6)\string: 1969--1972.

\bibitem{silvester2001efficient}
Silvester D, Elman H, Kay D, Wathen A. Efficient preconditioning of the
  linearized {N}avier--{S}tokes equations for incompressible flow. {\it J.
  Comput. Appl. Math.} 2001\string; 128(1-2)\string: 261--279.

\bibitem{kay2002preconditioner}
Kay D, Loghin D, Wathen A. A preconditioner for the steady-state
  {N}avier--{S}tokes equations. {\it SIAM J. Sci. Comput.} 2002\string;
  24(1)\string: 237--256.

\bibitem{fortin2000augmented}
Fortin M, Glowinski R. {\it Augmented {L}agrangian Methods: Applications to the
  Numerical Solution of Boundary-Value Problems}.
\newblock Elsevier .
\newblock 2000.

\bibitem{he2020efficient}
He X, Vuik C. Efficient and robust {S}chur complement approximations in the
  augmented {L}agrangian preconditioner for the incompressible laminar flows.
  {\it J. Comput. Phys.} 2020\string; 408\string: 109286.

\bibitem{nocedal2006numerical}
Nocedal J, Wright S. {\it Numerical Optimization}.
\newblock Springer Science \& Business Media .
\newblock 2006.

\bibitem{duff2001algorithms}
Duff IS, Koster J. On algorithms for permuting large entries to the diagonal of
  a sparse matrix. {\it SIAM J. Matrix Anal. Appl.} 2001\string; 22(4)\string:
  973--996.

\bibitem{amestoy1996approximate}
Amestoy PR, Davis TA, Duff IS. An approximate minimum degree ordering
  algorithm. {\it SIAM J. Matrix Anal. Appl.} 1996\string; 17(4)\string:
  886--905.

\bibitem{saad1988preconditioning}
Saad Y. Preconditioning techniques for nonsymmetric and indefinite linear
  systems. {\it J. Comput. Appl. Math.} 1988\string; 24(1-2)\string: 89--105.

\bibitem{yang2014scalable}
Yang C, Cai XC. A scalable fully implicit compressible {E}uler solver for
  mesoscale nonhydrostatic simulation of atmospheric flows. {\it SIAM J. Sci.
  Comput.} 2014\string; 36(5)\string: S23--S47.

\bibitem{saad1994ilut}
Saad Y. {ILUT}: {A} dual threshold incomplete {LU} factorization. {\it Numer.
  Linear Algebra Appl.} 1994\string; 1(4)\string: 387--402.

\bibitem{saad2005multilevel}
Saad Y. Multilevel {ILU} with reorderings for diagonal dominance. {\it SIAM J.
  Sci. Comput.} 2005\string; 27(3)\string: 1032--1057.

\bibitem{konshin2017lu}
Konshin I, Olshanskii M, Vassilevski Y. {LU} factorizations and {ILU}
  preconditioning for stabilized discretizations of incompressible
  {N}avier--{S}tokes equations. {\it Numer. Linear Algebra Appl.} 2017\string;
  24(3)\string: e2085.

\bibitem{mayer2007multilevel}
Mayer J. A multilevel {C}rout {ILU} preconditioner with pivoting and row
  permutation. {\it Numer. Linear Algebra Appl.} 2007\string; 14(10)\string:
  771--789.

\bibitem{bollhofer2006multilevel}
Bollh{\"o}fer M, Saad Y. Multilevel preconditioners constructed from
  inverse-based {ILU}s. {\it SIAM J. Sci. Comput.} 2006\string; 27(5)\string:
  1627--1650.

\bibitem{ghai2017comparison}
Ghai A, Lu C, Jiao X. A comparison of preconditioned {K}rylov subspace methods
  for large-scale nonsymmetric linear systems. {\it Numer. Linear Algebra
  Appl.} 2017\string; 26\string: e2215.

\bibitem{elman2008taxonomy}
Elman H, Howle VE, Shadid J, Shuttleworth R, Tuminaro R. A taxonomy and
  comparison of parallel block multi-level preconditioners for the
  incompressible {N}avier--{S}tokes equations. {\it J. Comput. Phy.}
  2008\string; 227(3)\string: 1790--1808.

\bibitem{briggs2000multigrid}
Briggs WL, Henson VE, McCormick SF. {\it A Multigrid Tutorial}. 72.
\newblock SIAM.
\newblock 2nd~ed. 2000.

\bibitem{lu2014hybrid}
Lu C, Jiao X, Missirlis N. A hybrid geometric+ algebraic multigrid method with
  semi-iterative smoothers. {\it Numer. Linear Algebra Appl.} 2014\string;
  21(2)\string: 221--238.

\bibitem{rudi2015extreme}
Rudi J, Malossi ACI, Isaac T, et al. An extreme-scale implicit solver for
  complex {PDE}s: highly heterogeneous flow in earth's mantle. In:  {\it
  Proceedings of the International Conference for High Performance Computing,
  Networking, Storage and Analysis}. ACM. ; 2015\string: 5.

\bibitem{elman2003parallel}
Elman HC, Howle VE, Shadid JN, Tuminaro RS. A parallel block multi-level
  preconditioner for the 3{D} incompressible {N}avier--{S}tokes equations. {\it
  J. Comput. Phy.} 2003\string; 187(2)\string: 504--523.

\bibitem{li2003crout}
Li N, Saad Y, Chow E. Crout versions of {ILU} for general sparse matrices. {\it
  SIAM J. Sci. Comput.} 2003\string; 25(2)\string: 716--728.

\bibitem{li2011supernodal}
Li XS, Shao M. A supernodal approach to incomplete {LU} factorization with
  partial pivoting. {\it ACM Trans. Math. Softw.} 2011\string; 37(4)\string:
  43.

\bibitem{jiao2020approximate}
Jiao X, Chen Q. Approximate generalized inverses with iterative refinement for
  $\epsilon$-accurate preconditioning of singular systems. {\it arxiv} 2020.
\newblock arXiv:2009.01673.

\bibitem{Golub13MC}
Golub GH, {Van Loan} CF. {\it Matrix Computations}.
\newblock Johns Hopkins.
\newblock 4th~ed. 2013.

\bibitem{dahl1992ilu}
Dahl O, Wille S. An {ILU} preconditioner with coupled node fill-in for
  iterative solution of the mixed finite element formulation of the 2{D} and
  3{D} {N}avier--{S}tokes equations. {\it Int. J. Numer. Methods Fluids}
  1992\string; 15(5)\string: 525--544.

\bibitem{saad1993flexible}
Saad Y. A flexible inner-outer preconditioned {GMRES} algorithm. {\it SIAM J.
  Sci. Comput.} 1993\string; 14(2)\string: 461--469.

\bibitem{amestoy2000mumps}
Amestoy PR, Duff IS, L{'}Excellent JY, Koster J. {MUMPS}: a general purpose
  distributed memory sparse solver. In:  {\it International Workshop on Applied
  Parallel Computing}. Springer. ; 2000\string: 121--130.

\bibitem{amestoy2019performance}
Amestoy P, Buttari A, L'Excellent JY, Mary T. Performance and Scalability of
  the Block Low-Rank Multifrontal Factorization on Multicore Architectures.
  {\it ACM Trans. Math. Software} 2019\string; 45\string: 2:1--2:26.

\bibitem{iluk2019miller}
Miller K. {ILU($k$)} preconditioner.
  \url{https://www.mathworks.com/matlabcentral/fileexchange/48320-ilu-k-preconditioner},
  {MATLAB} {F}ile {E}xchange;  2019.
\newblock Retrieved December 15, 2019.

\bibitem{boffi2013mixed}
Boffi D, Brezzi F, Fortin M, others . {\it Mixed {F}inite {E}lement {M}ethods
  and {A}pplications}. 44.
\newblock Springer .
\newblock 2013.

\bibitem{schafer1996benchmark}
Sch{\"a}fer M, Turek S, Durst F, Krause E, Rannacher R. Benchmark computations
  of laminar flow around a cylinder. In:  {\it Flow Simulation with
  High-Performance Computers II}. Springer.  1996 (pp. 547--566).

\bibitem{geuzaine2009gmsh}
Geuzaine C, Remacle JF. Gmsh: {A} 3-{D} finite element mesh generator with
  built-in pre-and post-processing facilities. {\it Int. J. Numer. Meth. Eng.}
  2009\string; 79(11)\string: 1309--1331.

\bibitem{bollhofer2006ilupack}
Bollh{\"o}fer M, Saad Y, Schenk O. {ILUPACK}-preconditioning software package.
  {\it Available online at the URL: http://ilupack.tu-bs.de} 2006.

\end{thebibliography}

\end{document}